\documentclass[pra,twocolumn,showpacs,superscriptaddress,nofootinbib]{revtex4-2}
\usepackage[table, svgnames, dvipsnames]{xcolor}
\usepackage{makecell, cellspace}
\usepackage{bbold}
\usepackage{amsmath, dsfont}
\usepackage{amsfonts}
\usepackage{bbm}
\usepackage{mathdesign}
\usepackage{mathtools}
\usepackage{amsthm}
\usepackage{bm}
\usepackage{amssymb}
\usepackage[normalem]{ulem}
\usepackage{braket}
\usepackage{hyperref}
\usepackage{tikz}
\usepackage{ifthen}
\usepackage{stmaryrd}
\usetikzlibrary{tikzmark}
\usepackage{upgreek}
\usepackage{braket}
\usepackage{physics}

\usepackage[]{lineno} 

\usepackage{graphicx}
\usepackage[shortlabels]{enumitem}

\usepackage{tikz-cd}
\usepackage{systeme}
\usepackage{bm}
\usepackage{tensor}
\usepackage{xparse}
\usepackage{amsmath}
\usepackage{physics}
\usepackage{amsthm}
\usepackage{mathtools}
\usepackage{amssymb}
\usepackage{dsfont}
\usepackage{xcolor}
\usepackage{upgreek}
\usepackage{mathrsfs}
\usepackage{quantikz}
\usepackage{tikz}
\usepackage{tikz-3dplot}
\usetikzlibrary{decorations.pathreplacing}
\definecolor{mycolor}{HTML}{90E0EF}
\usepackage{mathrsfs}
\usepackage{hyperref}

\usepackage{lettrine}
\usepackage{setspace}
\usepackage[utf8]{inputenc} 
\usepackage[T1]{fontenc}

\usepackage{wrapfig}

\theoremstyle{plain}

\theoremstyle{remark}

\DeclarePairedDelimiterX\hej[1]\lbrace\rbrace{#1}

\DeclarePairedDelimiterX{\braketHS}[2]{\langle}{\rangle_{\textsc{hs}}}{#1|#2}
\DeclarePairedDelimiterX{\expect}[1]{\langle}{\rangle}{#1}

\DeclarePairedDelimiterX{\inner}[2]{\langle}{\rangle}{#1, #2}
\DeclarePairedDelimiterX{\iner}[1]{\langle}{\rangle}{#1, #1}
\renewcommand{\selectlanguage}[1]{}

\usepackage{orcidlink}
\hypersetup{
	colorlinks=true,
	citecolor= BrickRed,
	linkcolor=BrickRed,
	filecolor=BrickRed,      
	urlcolor=BrickRed,
	pdftitle={Overleaf Example},
	pdfpagemode=FullScreen,
}

\usepackage{titlesec}
\titleformat{\section}
{\sffamily\bfseries} 
{\thesection} 
{1em} 
{} 

\titleformat{\subsection}
{\sffamily\bfseries} 
{\thesubsection} 
{1em} 
{} 

\usepackage[T1]{fontenc}
\usepackage[utf8]{inputenc}

\titlespacing\section{0pt}{12pt plus 4pt minus 2pt}{2pt plus 2pt minus 2pt}
\titlespacing\subsection{0pt}{12pt plus 4pt minus 2pt}{2pt plus 2pt minus 2pt}
\titlespacing\subsubsection{0pt}{12pt plus 4pt minus 2pt}{2pt plus 2pt minus 2pt}

\makeatletter
\let\@afterindenttrue\@afterindentfalse
\makeatother

\newcommand\blfootnote[1]{%
  \begingroup
  \renewcommand\thefootnote{}\footnote{#1}%
  \addtocounter{footnote}{-1}%
  \endgroup
}

\begin{document}
	\title{\textsf{Krylov complexity for non-local spin chains}}

	\author{Aranya Bhattacharya$^{\orcidlink{0000-0002-1882-4177}}$}
	\email{aranya.bhattacharya@uj.edu.pl}
	\affiliation{Institute of Physics, Jagiellonian University, Lojasiewicza 11, 30-348 Kraków, Poland. }
 
	\author{Pingal Pratyush Nath$^{\orcidlink{0000-0001-5311-7729}}$}
\email{pingalnath@iisc.ac.in}
	\affiliation{Centre for High Energy Physics, Indian Institute of Science, C.V. Raman Avenue, Bangalore 560012, India. }

	\author{Himanshu Sahu$^{\orcidlink{0000-0002-9522-6592}}$}
	\email{himanshusah1@iisc.ac.in}
     \affiliation{Centre for High Energy Physics, Indian Institute of Science, C.V. Raman Avenue, Bangalore 560012, India. }
	\affiliation{Department of Instrumentation \& Applied Physics, Indian Institute of Sciences, C.V. Raman Avenue, Bangalore 560012, Karnataka, India. }
	
\blfootnote{All authors contributed equally to this work. The authors of this paper were ordered alphabetically.}
	
	\begin{abstract}
		\noindent{\textbf{\textsf{Abstract}}}
		Building upon recent research in spin systems with non-local interactions, this study investigates operator growth using the Krylov complexity in different non-local versions of the Ising model. We find that the non-locality results in a faster scrambling of the operator to all sites. While the saturation value of Krylov complexity of local integrable and local chaotic theories differ by a significant margin, this difference is much suppressed when non-local terms are introduced in both regimes. This results from the faster scrambling of information in the presence of non-locality. In addition, we investigate the behavior of level statistics and spectral form factor as probes of quantum chaos to study the integrability breaking due to non-local interactions. Our numerics indicate that in the non-local case, late time saturation of Krylov complexity distinguishes between different underlying theories, while the early time complexity growth distinguishes different degrees of non-locality.\\
		
		\noindent DOI: \href{https://doi.org/10.1103/PhysRevD.109.066010}{10.1103/PhysRevD.109.066010}
	\end{abstract}
	
	\maketitle
	
	\vspace{1cm}
	
	\section{Introduction}
	\noindent The study of operator growth using Krylov complexity (also referred to as `K-complexity') has emerged as a powerful framework for investigating the behavior of quantum systems and their underlying dynamics\,\cite{PhysRevX.9.041017,barbonEvolutionOperatorComplexity2019a,rabinoviciOperatorComplexityJourney2021b}. K-complexity is a measure of the delocalization of a local initial operator evolving under Heisenberg evolution with respect to the Hamiltonian \cite{PhysRevX.9.041017,barbonEvolutionOperatorComplexity2019a,rabinoviciOperatorComplexityJourney2021b,PhysRevB.102.085137,jianComplexityGrowthOperators2021}. It is conjectured to grow at most exponentially generically non-integrable systems\,\cite{PhysRevX.9.041017}. This exponential growth of K-complexity can be used to extract the Lyapunov exponent\,\cite{PhysRevX.9.041017,barbonEvolutionOperatorComplexity2019a}, establishing a connection with out-of-time-ordered-correlators (OTOC)\,\cite{maldacenaBoundChaos2016,PhysRevResearch.2.043234}. Further studies have shown a relation between K-complexity and chaos in context of various models such as Ising models\,\cite{rabinovici_krylov_2022,rabinoviciKrylovLocalizationSuppression2022}, Sachdev-Ye-Kitaev (SYK) models \cite{jianComplexityGrowthOperators2021,bhattacharjeeKrylovComplexityLarge2023,heQuantumChaosScrambling2022}, quantum field theories\,\cite{caputaOperatorGrowth2d2021,khetrapalChaosOperatorGrowth2023,kunduStateDependenceKrylov2023,camargoKrylovComplexityFree2023,avdoshkinKrylovComplexityQuantum2022, Erdmenger:2023wjg}, many-body localization system\,\cite{10.21468/SciPostPhys.13.2.037, Bento:2023bjn}, and open quantum systems\,\cite{bhattacharyaOperatorGrowthKrylov2022b,bhattacharjeeOperatorGrowthOpen2023,bhattacharyaKrylovComplexityOpen2023d,PhysRevResearch.5.033085, Bhattacharjee:2023uwx}.

Among different systems, of particular interest are the so-called fast scramblers for which scrambling time is logarithmic in the number of degrees of freedom. The examples include black holes\,\cite{Yasuhiro_Sekino_2008,lashkariFastScramblingConjecture2013,maldacenaBoundChaos2016}, conjectured to be the fastest scramblers in nature, SYK\,\cite{PhysRevLett.70.3339} and other related holographic models. The previous studies\,\cite{PhysRevA.94.040302,PhysRevA.99.051803,PhysRevLett.123.130601,belyansky2020minimal} suggest that local chaotic dynamics along with non-local interactions are sufficient to give rise to the phenomena termed as ``fast scrambling". These developments have motivated the study of quantum information scrambling in non-local systems\,\cite{PhysRevResearch.2.043399,wanisch2022quantum} as well as experimental proposal for probing fast scramblers with simpler models\,\cite{PhysRevLett.123.130601,belyansky2020minimal,PhysRevResearch.2.043399,PhysRevB.107.L081103}.\\
	
	\noindent In this study, we characterize the growth of K-complexity in non-local systems. It has been established that K-complexity can distinguish between the integrable and chaotic regimes in local systems. To be precise, the complexity growth rate and saturation value are known to be significantly higher for chaotic systems as compared to the integrable ones\,\cite{rabinoviciOperatorComplexityJourney2021b,rabinovici_krylov_2022,rabinoviciKrylovLocalizationSuppression2022}. Here, we ask the question of how these characteristics get modified once non-local interactions are turned on within these regimes. While chaotic systems with non-local interactions can exhibit fast scrambling properties, the non-local terms in the otherwise integrable cases are also expected to accelerate the operator growth and therefore begs some investigation as to whether the notions of local integrability and chaos remain distinguishable in the presence of non-locality.  

Apart from K-Complexity, we also study the two widely used measures to probe chaos: Level Statistics (Appendix \ref{subsec:level_statistics}) and Spectral Form Factor (Appendix \ref{subsec:SFF}). The symmetry-reduced level statistics is known to exhibit Poisson distribution for integrable regimes, while in chaotic regimes, it shows the Wigner-Dyson distribution\,\cite{10.1007/978-3-0348-8266-8_36,PhysRevLett.52.1,PhysRevLett.110.084101}. On the other hand, the spectral form factor (SFF) is found to exhibit an explicit dip-ramp-plateau behaviour for chaotic evolution in contrast to the integrable regime where higher amount of fluctuations wash out the dip-ramp-plateau behaviour\,\cite{cotlerBlackHolesRandom2017,PhysRevD.95.126008,winerSpectralFormFactor2022}.\footnote{This might vary from model to model. In some cases, an effective ramp appears, but it is not as evidently clear as it is for chaotic ones.} We study these quantities to compare their behaviour in the presence of non-local terms in the Hamiltonian and analyse how the introduction of non-locality changes their nature.

Our results indicate that with increasing degrees of non-locality, the integrable evolution becomes similar to chaotic ones. We study the three diagnostics for characterizing chaos for various models which indicates that non locality apparently makes all regimes chaotic. However, a closer look at the plots reveals that non-local terms with local integrable operator evolution can still be distinguished from non-local terms with local chaotic operator evolution using  K-complexity. Moreover, we also observe that the increase in non-locality is captured by the initial growth rate of the K-complexity. To have a clearer understanding of how non-locality changes behavior of Krylov complexity, we also study ``mixed field" spin chains\,\cite{wanisch2022quantum}, where we vary the degree of non-locality and study how it increases the slope of the Lanczos ascent.
	
The rest of the paper is structured as follows. In section \ref{sec:review_for_local_hamiltonians}, we briefly review the notion of K-complexity, and it's characteristics for local Hamiltonians, discussing both integrable and non-integrable regimes. In section \ref{sec:Non_local_results}, we introduce several non-local models and discuss distinguished features present in the K-complexity profiles. Section \ref{subsec:Non-local Transverse Field Ising Model} consists of the numerical findings for non-locality introduced to an otherwise local integrable and chaotic version of the transverse field Ising model. In sections \ref{subsec:Transverse Mixed Field Ising} and \ref{subsec:mixedfieldXXZ}, we present the models and numerics when we deal with varying degrees of non-locality by tuning a parameter in transverse field Ising model and XXZ spin chain respectively. Finally, we conclude in section \ref{sec:Discussion} discussing the implications of our results. In Appendix we study the remaining probes, namely the level statistics and the spectral form Factor for both local and non-local models to be able to compare their behavior with Krylov complexity.

	\section{Review for local Hamiltonians}\label{sec:review_for_local_hamiltonians}
	
In this section, we briefly review Krylov complexity and the features of operator growth for local Hamiltonians. As part of our exploration, we will focus on how this behaves within the context of the transverse-field Ising model, considering both integrable and non-integrable scenarios. For our analysis, we will take the local Hamiltonian to be a one-dimensional transverse-field Ising model with open boundary conditions, which has the form, 
\begin{align}
    H_{\mathrm{TFIM}} = - \sum_{j=1}^{N-1}  \sigma^{z}_{j} \sigma^{z}_{j+1} - g \sum_{j=1}^{N} \sigma^{x}_j - h \sum_{j=1}^{N} \sigma^{z}_{j}\,, \label{tfim}
\end{align}
where $g$ and $h$ are the coupling parameters. When $h=0$, the Hamiltonian remains integrable for all values of $g$, as it can be mapped to the free-fermionic model \cite{sachdev_2011}. However, when both $g$ and $h$ are nonzero, the system departs from integrability. To investigate the integrable regime, we set $g=1$ and $h=0$, while for the chaotic regime, we choose $g=-1.05$ and $h=0.5$ \cite{PhysRevLett.106.050405}. The model has a parity symmetry for all values of coupling parameters, and additional $Z$-reflection symmetry when $h = 0$.

	\subsection{Operator dynamics in Krylov basis}\label{subsec:operator_dynamics_in_krylov_basis}
	
	We begin with a brief review of the operator dynamics in Krylov space\cite{PhysRevX.9.041017, rabinoviciOperatorComplexityJourney2021b, barbonEvolutionOperatorComplexity2019a}. Consider the evolution of an operator with seed $\mathcal{O}_0$ at $t =0$ under the time-independent Hamiltonian $H$. At any time $t$, the time-evolved operator $\mathcal{O}(t)$  under Heisenberg evolution can be written as
	\begin{equation}
		\mathcal{O}(t) = e^{iHt}\mathcal{O}_0e^{-iHt} = e^{i\mathcal{L}t}\mathcal{O}_0 = \sum_{n=0}^\infty \frac{(it)^n}{n!}\mathcal{L}^n\mathcal{O}_0
	\end{equation}
	where $\mathcal{L}$ is Liouvillian superoperator, which satisfies the relation $\mathcal{L}\mathcal{O}=[H,\mathcal{O}]$. This motivates the definition of Krylov space associated with the operator $\mathcal{O}$ as the minimal subspace of operator space that contains the time evolution of $\mathcal{O}$ at all times. Therefore, the Krylov space is obtained from repeated action of Liouvillian $\mathcal{L}^n\mathcal{O}$ :
	\begin{equation}
		\begin{split}
				\mathcal{H}_\mathcal{O} &= \text{span}\{ \mathcal{L}^n \mathcal{O}\}^\infty_{n=0} \\
				&= \text{span} \{\mathcal{O},[H,\mathcal{O}],[H,[H,\mathcal{O}]],\ldots  \}
		\end{split}
	\end{equation}
Once the Krylov space is obtained, an orthonormal basis is then formed using a choice of inner product on the operator space. This is achieved using the Lanczos algorithm, an instance of the Gram-Schmidt process\,\cite{PhysRevX.9.041017,baiTemplatesSolutionAlgebraic1987}. For our purpose, we will use the infinite-temperature inner product, also known as Frobenius inner product :
\begin{equation}
    (\mathcal{A}|\mathcal{B}) = \frac{1}{D} \text{Tr}\left[\mathcal{A}^\dagger \mathcal{B}\right],\ \ \ \lVert \mathcal{A} \rVert = \sqrt{(\mathcal{A}|\mathcal{A})},
\end{equation}
	
 where $D=\text{Tr}[\textbf{I}]$ is the trace of the identity matrix of appropriate dimension. In the orthonormal basis obtained from Lanczos iteration, the Liouvillian takes a tridiagonal form :

\begin{equation}\label{eq:liov_tri}
    \mathcal{L} = \begin{pmatrix}0&b_1&&&&0\\ b_1&0&b_2 &&&\\&b_2&0&\ddots &&\\&&\ddots &\ddots &b_{K-2}&\\&&&b_{K-2}&0&b_{K-1} \\0&&&&b_{K-1}&0\\\end{pmatrix}
\end{equation}
where $K$ is dimension of Krylov space, constrained to $1\leq K\leq D^2 - D + 1$ \cite{rabinoviciOperatorComplexityJourney2021b}. Eq. \eqref{eq:liov_tri} can be recast into the following form, 
\begin{equation}
    \mathcal{L}|\mathcal{O}_n ) = b_n |\mathcal{O}_{n-1}) + b_{n+1}|\mathcal{O}_{n+1})
\end{equation}
We can expand the time-evolving operator in the Krylov basis as :
\begin{equation}
    |\mathcal{O}(t) ) = \sum_{n=0}^{K-1}i^n \phi_n(t) |\mathcal{O}_n )
\end{equation}
where $\phi_n(t)$ are time-dependent probability amplitudes associated with the Krylov chain. The wavefunctions satisfy the recursive differential equation, followed by the Heisenberg equation, which takes the form
\begin{equation}\label{eq:reccu}
    \dot{\phi}_n(t) = b_n \phi_{n-1}(t) - b_{n+1}\phi_{n+1}(t)
\end{equation}
where $\phi_{-1}(t) = 0$ and $\phi_n(0) = \delta_{n0}$. The unitarity of operator evolution implies a normalization condition on the probability amplitudes,
\begin{equation}
    \sum_{n = 0}^{K-1} |\phi_n(t) |^2  = 1
\end{equation}

	In Eq.~\eqref{eq:reccu}, we can think of the Lanczos coefficients $b_n$ as the hopping amplitudes for the seed $O_0$ to traverse in the Krylov Space and $\phi_n$ as the probability amplitude associated with site $n$. The average position of the distribution on the Krylov chain is called K-complexity and is given by
	\begin{equation}
		C_K(t) = \sum_{n=0}^{K-1} n |\phi_n(t) |^2
	\end{equation}
	We can also define the complexity operator\,\cite{hornedal_ultimate_2022}, 
	\begin{equation}
		\mathcal{K} = \sum_{n=0}^{K-1} n |\mathcal{O}_n ) ( \mathcal{O}_n | 
	\end{equation}
	which plays the role of the position operator in the Krylov chain so that the Krylov complexity can be written as the expectation value of the complexity operator $\mathcal{K}$
	\begin{equation}
		C_K(t) = (\mathcal{O}(t) |\mathcal{K}|\mathcal{O}(t))\,.
	\end{equation}
	
	\begin{figure}[h]
		\centering
			\includegraphics[scale = 0.29]{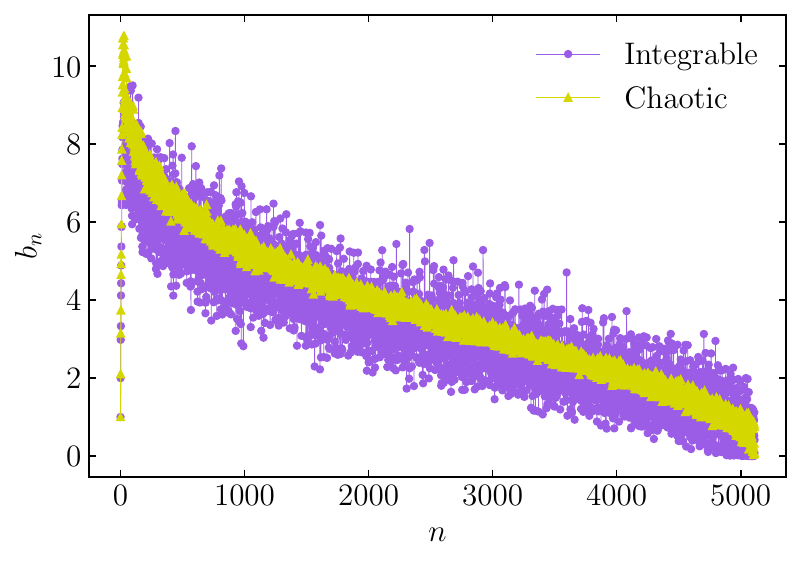}
			\includegraphics[scale = 0.29]{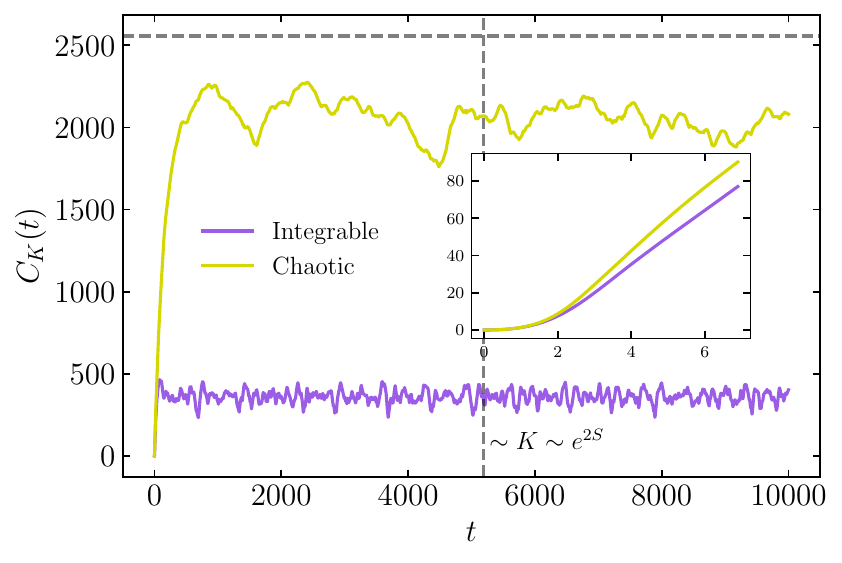}
		\caption{\textbf{Left:} The Lanczos-sequence for the TFIM Hamiltonian in Eq.~\eqref{tfim} computed for $L = 7$ spins with open boundary conditions for fixed parity $P = +1$ in (a) Integrable (b) Chaotic limits. \textbf{Right:} Krylov complexity for TFIM Hamiltonian \eqref{tfim} computed for $L = 7$ spins with open boundary conditions for fixed parity $P = +1$ in Integrable and Chaotic limits. \textbf{Outset:} Full-time range computed. \textbf{Inset:} Zoom in at early times.}
		\label{fig:local_Krylov}
	\end{figure}
	To study K-complexity we will use open boundary conditions and focus on a local operator $\mathcal{O}$ which respects the parity symmetry and keeps the computation within the chosen sector,
	
	\begin{equation}\label{eq:in_op}
		\mathcal{O} = S^z_i + S^z_{N-i+1},
	\end{equation}
where $i$ is chosen somewhere close to the center of the chain.\footnote{The operator is chosen so that it respects the parity symmetry (defined as in \cite{10.1119/1.4798343}) and keeps the computation within the chosen sector. The results deduced in this work remains unchanged even if $i$ is varied in Eq.~\eqref{eq:in_op}. However, as discussed in \cite{rabinovici_krylov_2022}, the behavior of K-complexity is controlled by the statistics of the Hamiltonian spectrum and the structure of operator under consideration. For example, if the chosen operator has spread in all the sites from the beginning, the behaviour of Krylov complexity might change as there is no scope for scrambling. Therefore, while the results can differ for an arbitrary seed operator, the conclusions remain invariant for any operator that respects the symmetries (parity here) of the Hamiltonian and has support in small number of sites to begin with so that it has enough scope to spread and scramble in the rest of the sites.} Fig.~\ref{fig:local_Krylov} shows the Lanczos sequence $b_n$ and  K-complexity  for TFIM Hamiltonian \eqref{tfim} computed for $L = 7$ for fixed parity $P =+1$ sector for the operator $\mathcal{O} = S^z_4$. The study of Lanczos sequence and Krylov complexity in integrable and chaotic limits have extensively been done in previous studies for different systems\,\cite{rabinovici_krylov_2022,iizukaKrylovComplexityIP2023,bhattacharyyaOperatorGrowthKrylov2023,camargoSpectralKrylovComplexity2023}. The Lanczos coefficient features sublinear growth in the integrable limit while linear growth in the chaotic limit, followed by saturation, and the descent, while  K-complexity transitions from exponential growth at very early time to linear increase followed by saturation\,\cite{rabinoviciOperatorComplexityJourney2021b,rabinovici_krylov_2022}. The saturation value of K-complexity is significantly large for local chaotic evolution as compared to the local integrable one. The suppressed complexity saturation of integrable cases results from higher fluctuations in Lanczos coefficients during the descent phase. This fluctuating descent period is dubbed as the Krylov localization.

\section{Non-local models and numerical findings:}\label{sec:Non_local_results}
In this section, we will study operator growth in non-local systems. We compare the results with the local counterpart to demonstrate the effect of non-locality on these quantities. For these Hamiltonians, it is, in general, expected that due to non-local terms, the Liouvillian will also be non-local. Now, since the notion of integrability is believed to be closely related to locality \cite{PhysRevResearch.2.043399,wanisch2022quantum}, the evolution of the operator under the non-local Liouvillian is expected to show chaotic behavior even in the case when the local part of the Hamiltonian is integrable. Therefore it is expected that the non-local terms in the Hamiltonian will result in the loss of its integrability.

	We study two types of non-local terms in this paper: i) All-to-all couplings with the same interaction strength which we call the non-local transverse field Ising model, and ii) All-to-all couplings with varying interaction strength, where the coupling between two sites decay as a power of the distance between the sites. We call this the transverse mixed field Ising model and by tuning the power, we vary the degree of non-locality in this model.

	\subsection{Non-local Transverse Field Ising Model}\label{subsec:Non-local Transverse Field Ising Model}

We consider the fast scrambling spin-1/2 Hamiltonian introduced in \cite{belyansky2020minimal},
\begin{equation}\label{eq:Ham_fast}
    \mathcal{H} = \mathcal{H}_\text{local} - \frac{\gamma}{\sqrt{N}} \sum_{i<j} \sigma^z_i \sigma^z_j,
\end{equation}
where $\sigma^z_i$ is the Pauli $z$ operator acting on site $i$ and $\mathcal{H}_\text{local}$ is Hamiltonian with only local interactions. In our study, we will consider the local Hamiltonian as a  one-dimensional transverse-field Ising model with open boundary conditions introduced in Sec.~\ref{sec:review_for_local_hamiltonians}. At this point, it is important to emphasize the terminology used throughout this literature. We will later observe from the results of operator growth as well as level-statistics and spectral form factor (discussed in the appendices \ref{subsec:level_statistics} and \ref{subsec:SFF}), that non-locality can break the integrability of the system. While referring to a system whose local part of the Hamiltonian in Eq.~\eqref{eq:Ham_fast} is integrable, we will call it non-local integrable Hamiltonian even though non-locality breaks integrability, and the model is no longer integrable.\\

	\begin{figure}[h]
		\centering

			\includegraphics[scale = 0.29]{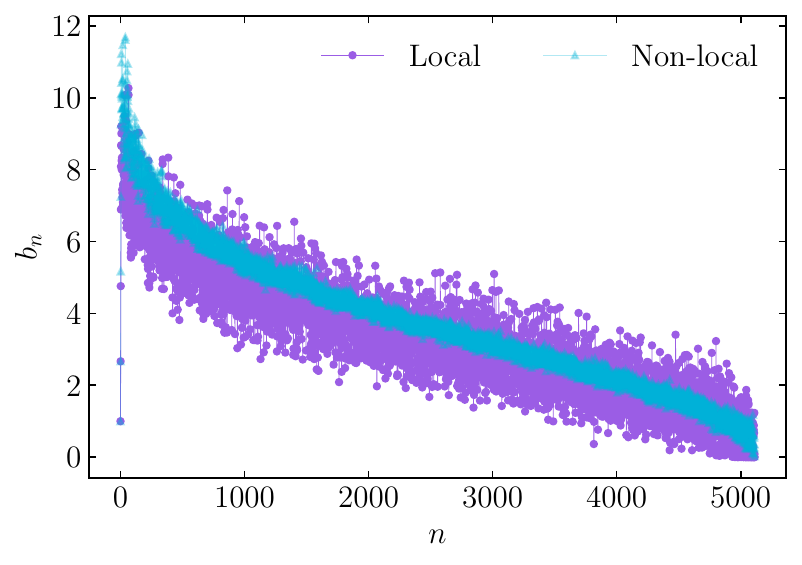}
			\includegraphics[scale = 0.29]{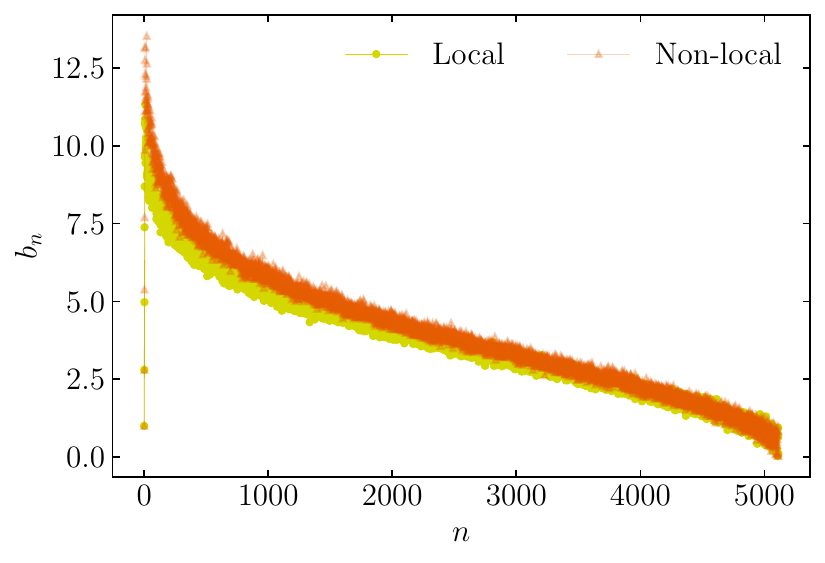}
			\includegraphics[scale = 0.29]{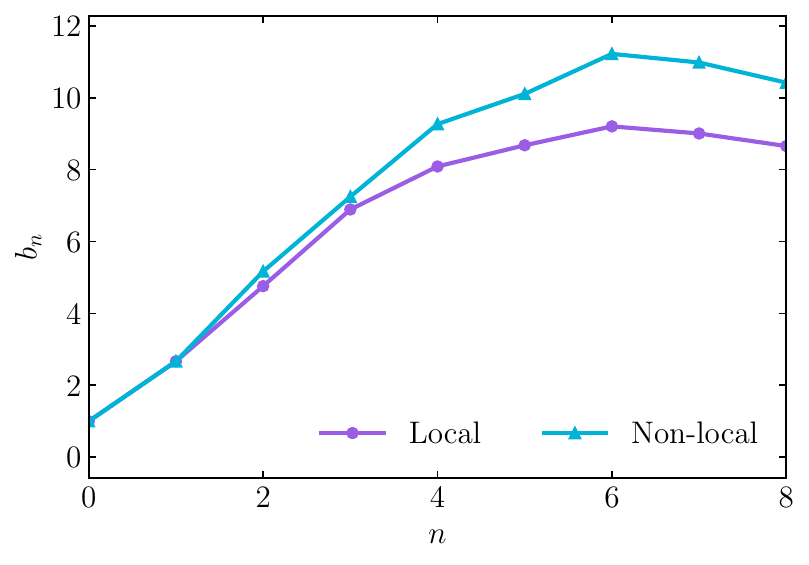}
			\includegraphics[scale = 0.29]{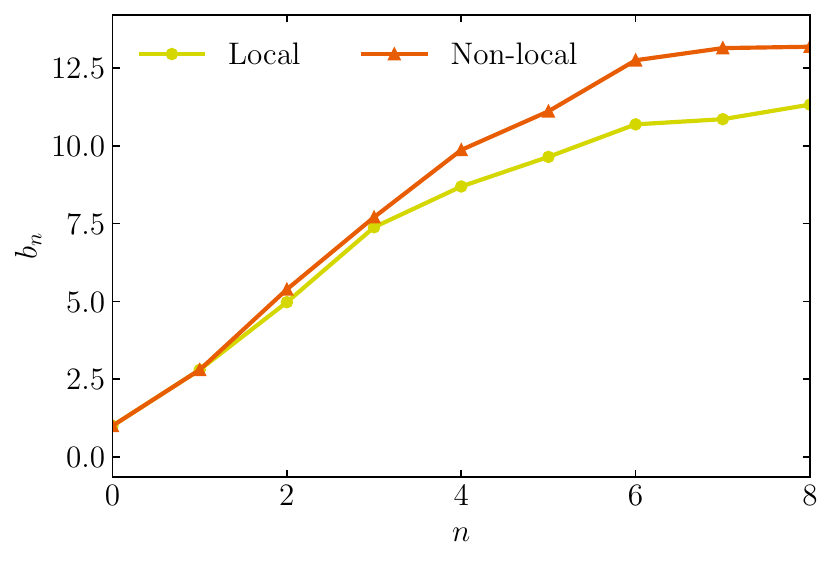}

		\caption{Comparison between Lanczos-coefficients for local and non-local TFIM Hamiltonian with non-local parameter $\gamma = 0.5$, computed for  $L=7$ spins in the $P = +1$ sector for Integrable and Chaotic limits of local Hamiltonian. \textbf{Top row}: complete spectrum of Lanczos coefficients, \textbf{Bottom row}: Initial lanczos coefficients.}
		\label{fig:lanczos_NL}
	\end{figure}

	\begin{figure}[h]
		\centering
			\includegraphics[scale = 0.29]{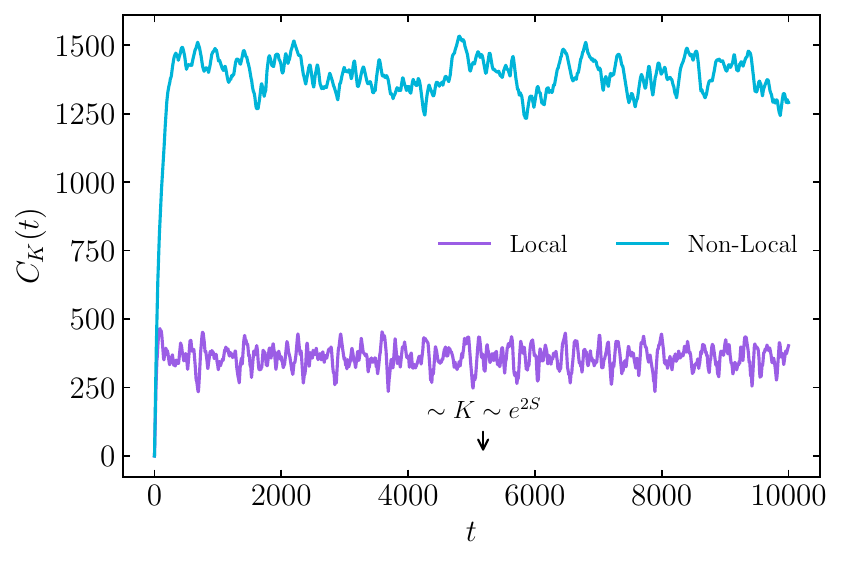}
			\includegraphics[scale = 0.29]{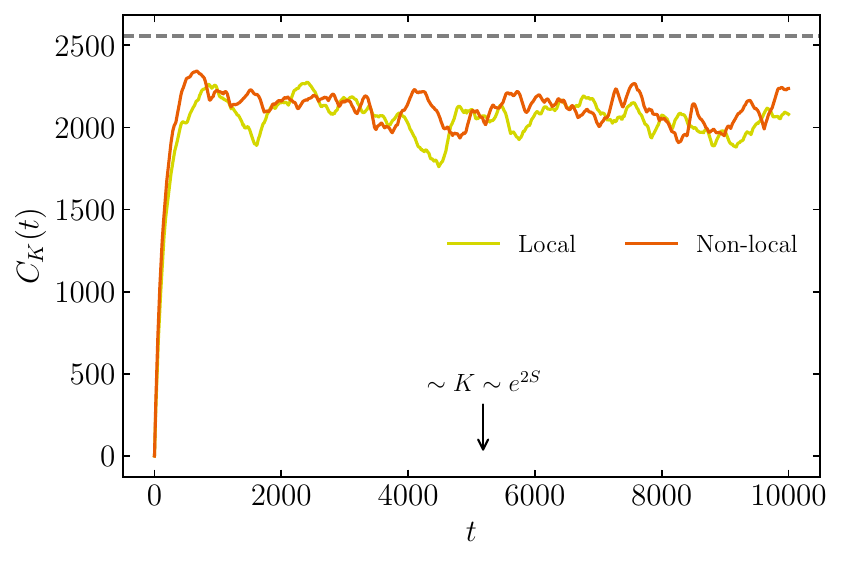}
			\includegraphics[scale = 0.29]{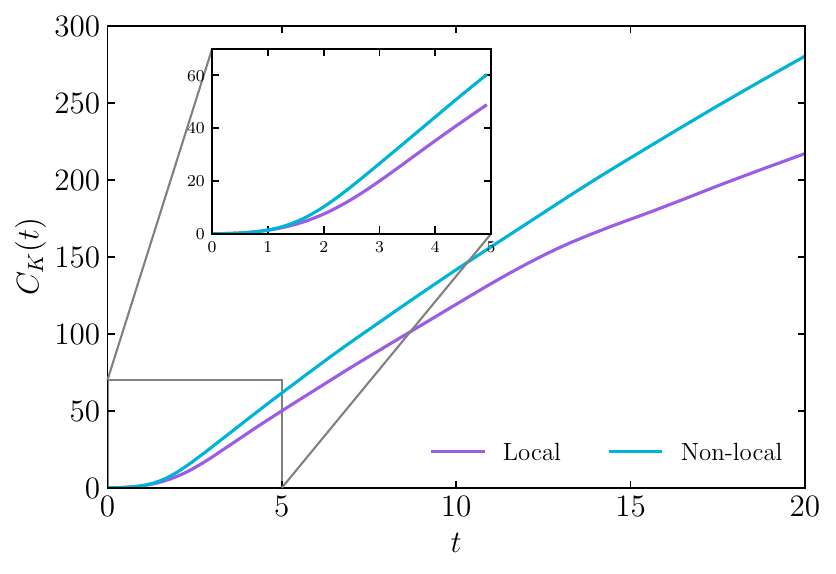}
			\includegraphics[scale = 0.29]{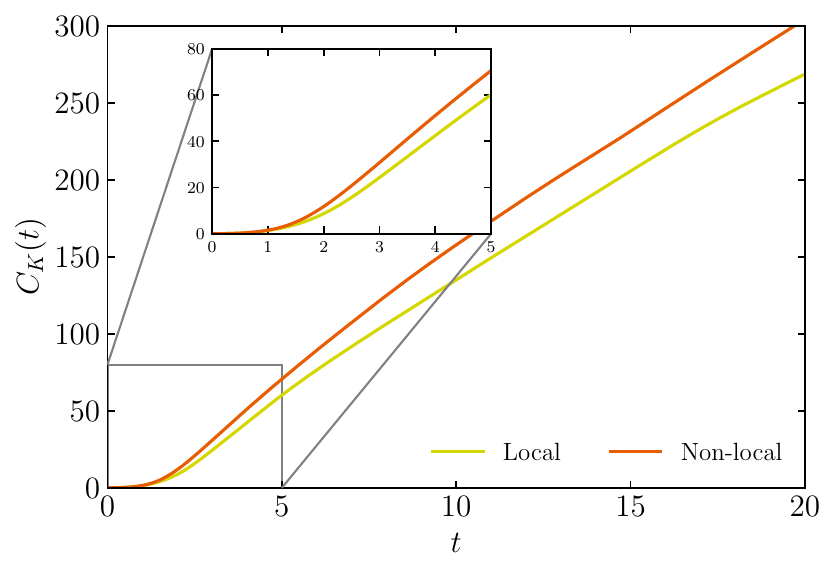}
		\caption{Krylov complexity for TFIM Hamiltonian \eqref{tfim} computed for $L = 7$ spins with open boundary conditions for fixed parity $P = +1$, and non-locality parameter $\gamma = 0.5$ in \textbf{Left:} Integrable \textbf{Right:} Chaotic limits. \textbf{Top:} Full-time range computed. \textbf{Bottom:} Zoom in at early times.}
		\label{fig:non-local_complexity}
	\end{figure}

 \begin{figure}[h]
     \centering
     \includegraphics[width = 0.48\linewidth]{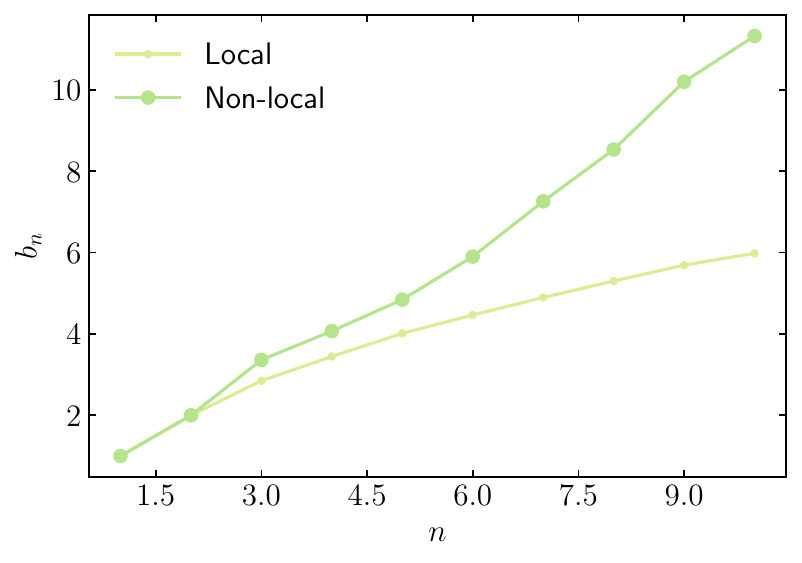}
     \includegraphics[width=0.48\linewidth]{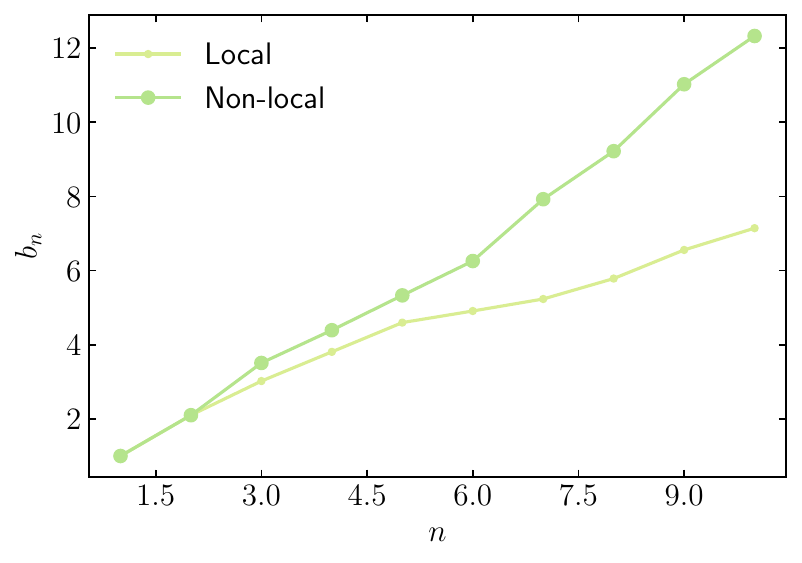}
     \caption{Comparison between initial Lanczos-coefficients for local and non-local TFIM Hamiltonian with non-local parameter $\gamma = 0.5$, computed for  $L=12$ spins in the $P = +1$ sector for Integrable and Chaotic limits of local Hamiltonian. The Lanczos coefficients exhibit faster growth with non-local terms, especially evident in larger system sizes.}
     \label{fig:FS_Initial}
 \end{figure}

\noindent  We studied the non-local TFIM Hamiltonian with $L=7$ sites and non-local parameter $\gamma = 0.5$. The numerical results for Lanczos-sequence and Krylov complexity are shown in Fig.~\ref{fig:lanczos_NL} and \ref{fig:non-local_complexity}. The global picture emerging from these results can be summarized in the following points:
	
	\begin{itemize}
    \item \textbf{\textsf{Lanczos sequence}} The initial growth of Lanczos coefficients is faster for the non-local cases as compared to local cases (in both integrable and chaotic limits). While this accelerated growth is not prominently observable in smaller system sizes, it becomes distinctly apparent in larger system sizes as shown in Figure \ref{fig:FS_Initial}. Note that the growth of the primary few Lanczos coefficients correspond to the pre-scrambling behaviour. Hence a faster growth of primary Lanczos coefficients result in an higher exponent for the exponentially growing complexity before scrambling time. The peak value of the Lanczos sequence is more than that of the corresponding local case. The overall Lanczos profile for the non-local integrable case resembles the form of the local chaotic case. The  fluctuations in the decaying part of the Lanczos sequence are also small for the  non-local integrable case which results in a high saturation value of K-complexity.
    \item \textbf{\textsf{K-complexity}} This shows a transition from exponential growth to linear increase starting from the scrambling time. This transition time is lower for non-local models in both integrable and chaotic regimes. Furthermore, the initial growth rate is also greater in non-local cases compared to their local counterparts. At exponentially late times the complexity saturates at half of the Krylov dimension for the chaotic cases local and non-local cases, since by then the operator is uniformly distributed over the Krylov basis. It is also worth noting that even after the introduction of non-local terms to the otherwise local integrable Hamiltonian, the complexity saturation value does not reach exactly the value of the local chaotic Hamiltonian. The saturation value for non-local terms $+$ local integrable Hamiltonian is close to $1500$ (much higher compared to local integrable which was $500$) whereas the the saturation value for non-local terms $+$ local chaotic Hamiltonian is close to $2000$ (almost similar to the local chaotic case). The late time saturation value of the chaotic Hamiltonian is the same for both local and non-local cases, although the non-local chaotic Hamiltonian has a large initial growth compared to the local chaotic Hamiltonian.    
\end{itemize}

	\subsection{Transverse Mixed Field Ising Model}\label{subsec:Transverse Mixed Field Ising}
Consider the one-dimensional transverse mixed field Ising model of $N$ spins with open boundary conditions 
\begin{equation}\label{eq:mixed_TFIM}
    \mathcal{H}_\alpha = -\sum_{i<j} J^\alpha_{ij} \sigma^z_i \sigma^z_j - g \sum_{i} \sigma^x_j- h \sum_{j} \sigma^{z}_{j}
\end{equation}
where $J^\alpha_{ij}$ is interaction strength between spins at position $i$ and $j$ which assume to follow powerlaw $1/\kappa \cdot J/|i-j|^\alpha$. Throughout our study, we take $J=1$, and $\kappa =1$. In previous studies \cite{wanisch2022quantum}, the information scrambling, using out-of-time-order correlators (OTOCs) as a probe, has been studied in such a model with a variation of non-local parameter $\alpha$. It's shown that for $\alpha > 2$, the dynamics effectively resemble local dynamics, while for smaller exponents, the dynamics become non-local. Here, we will confine ourselves to studying the initial Lanczos sequence for the model with varying exponent $\alpha$.\\

	\begin{figure}[h]
		\centering
			\includegraphics[scale=0.29]{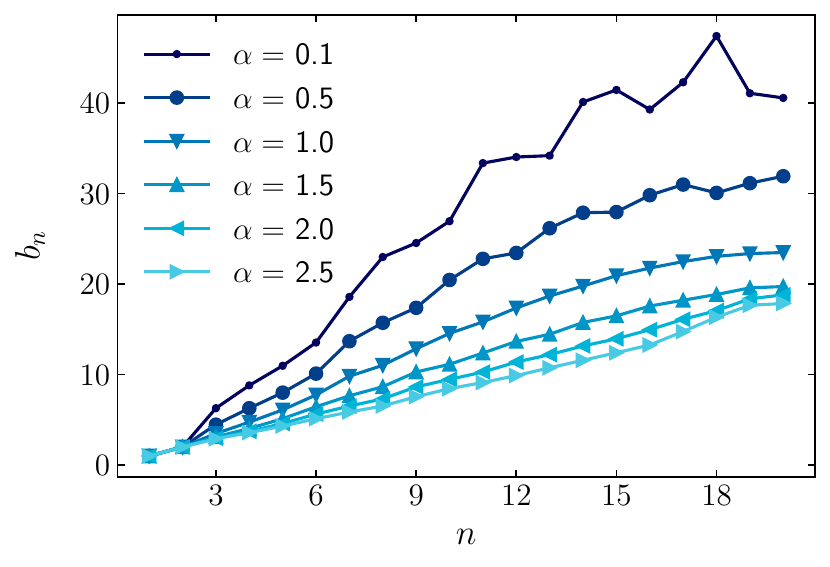}
			\includegraphics[scale=0.29]{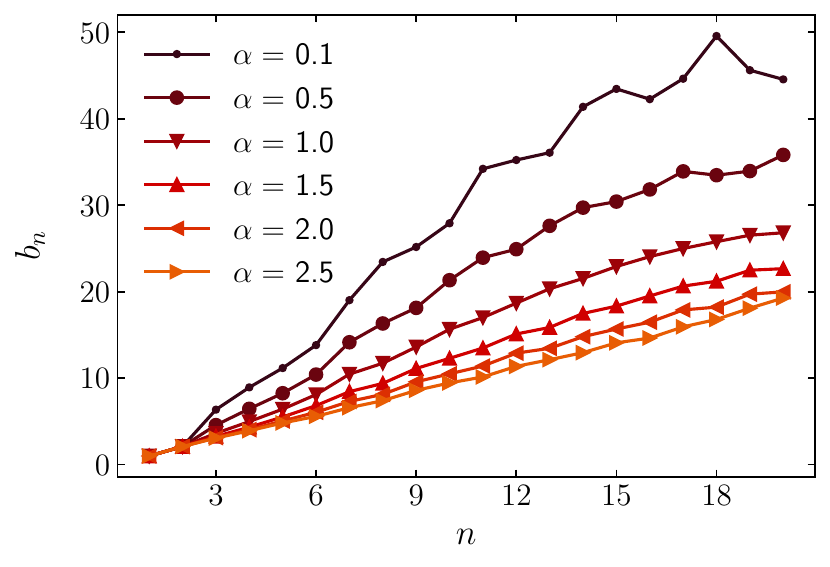}
		\caption{The Lanczos-sequence for the Transverse Mixed Field Ising Model Hamiltonian in Eq.~\eqref{eq:mixed_TFIM} computed for $L = 13$ spins with open boundary conditions for fixed parity $P = +1$ in (a) Integrable (b) Chaotic limits with varying non-local exponent $\alpha$ values. }
		\label{fig:lanczos_mixed_TFIM}
	\end{figure}

	\begin{figure}[h]
		\centering
			\includegraphics[scale=0.29]{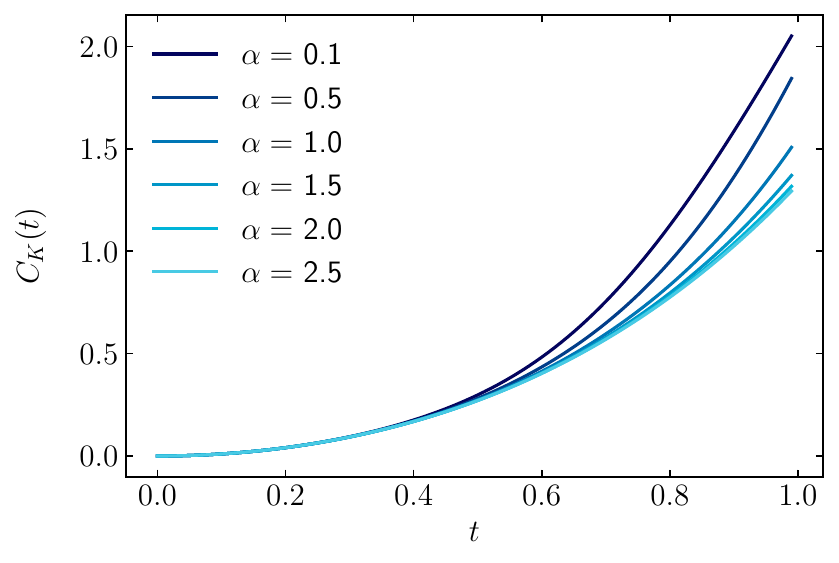}
			\includegraphics[scale=0.29]{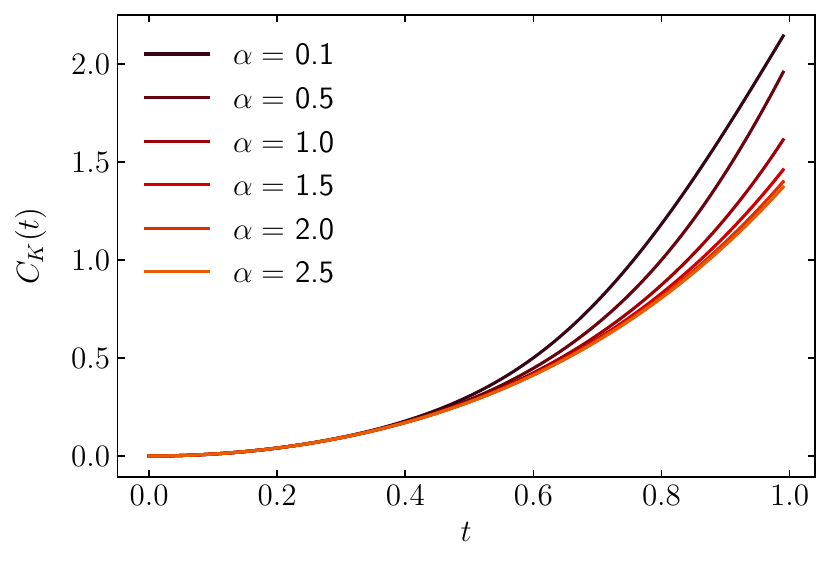}
		\caption{Krylov complexity for the Transverse Mixed Field Ising Model Hamiltonian in Eq.~\eqref{eq:mixed_TFIM} computed for $L = 13$ spins with open boundary conditions for fixed parity $P = +1$ in (a) Integrable (b) Chaotic limits with varying non-local exponent $\alpha$ values. }
		\label{fig:complexity_mixed_TFIM}
	\end{figure}
	
\begin{figure}[h]
		\centering
			\includegraphics[scale=0.29]{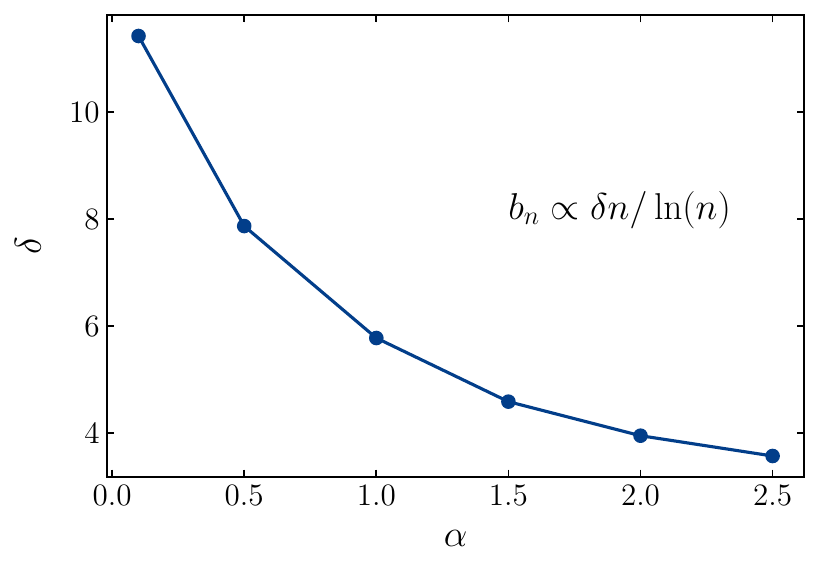}
			\includegraphics[scale=0.29]{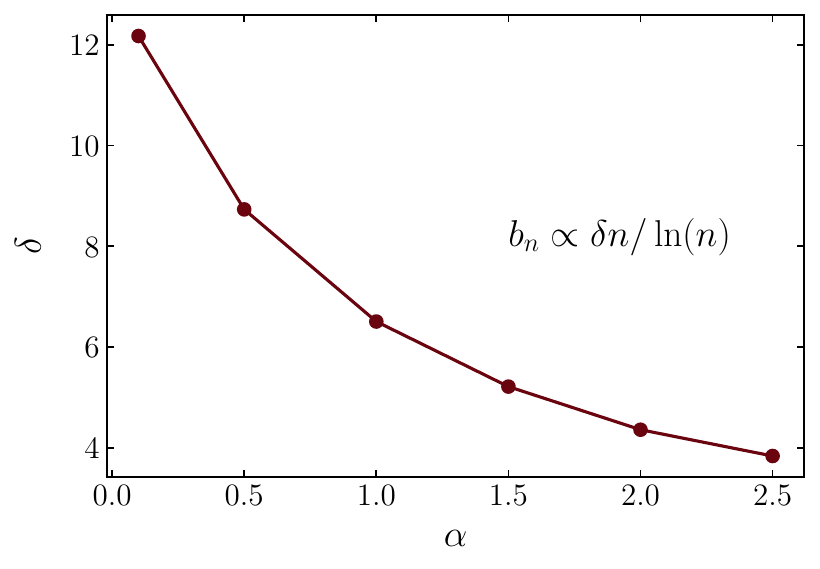}
		\caption{The growth rate $\delta$ plotted for different non-local exponent $\alpha$ values for Lanczos coefficients in Fig.~\ref{fig:lanczos_mixed_TFIM}. }
		\label{fig:slope_mixed_TFIM}
	\end{figure}

\noindent In Fig.~\ref{fig:lanczos_mixed_TFIM}, we showed the initial Lanczos coefficients $b_n$ for $L = 13$ with varying exponent $\alpha \in \{0.1,0.5,1.0,1.5,2.0,2.5\}$. The initial operator is chosen to be $S^z_7$. In both limits integrable and chaotic, we find the increasing slope and saturation value in initial growth with the decrease in exponent $\alpha$ and therefore increase in non-locality. This ensures that the initial (pre-scrambling) growth rate of the Krylov complexity always increases in the presence of non-locality (See Fig.~\ref{fig:complexity_mixed_TFIM}). In Figure \ref{fig:slope_mixed_TFIM}, we showed the initial growth rate calculated using fit ,$b_n = \delta \frac{n}{ln[n]} + c$ (for $1$D systems, as conjectured in \cite{PhysRevX.9.041017}) in initial Lanczos coefficients. 

	\begin{figure}[h]
     \centering
     \includegraphics[width = 0.48\linewidth]{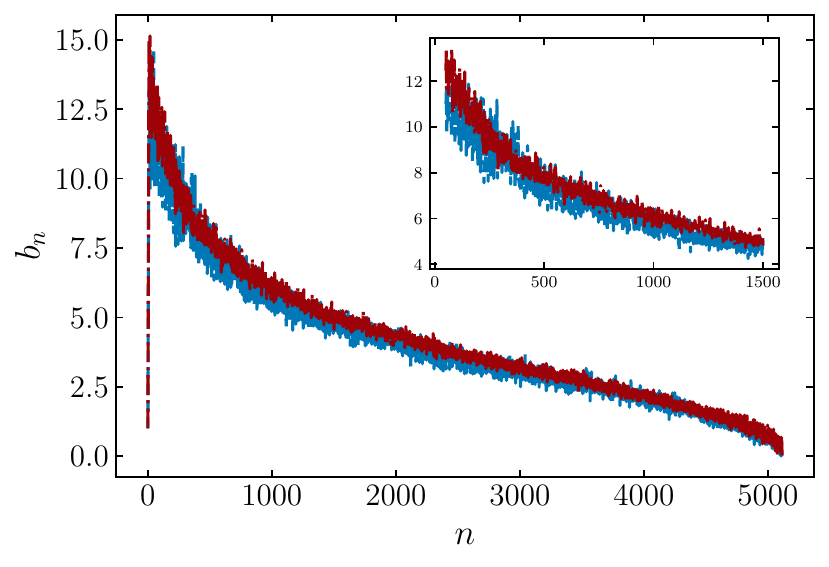}
     \includegraphics[width=0.48\linewidth]{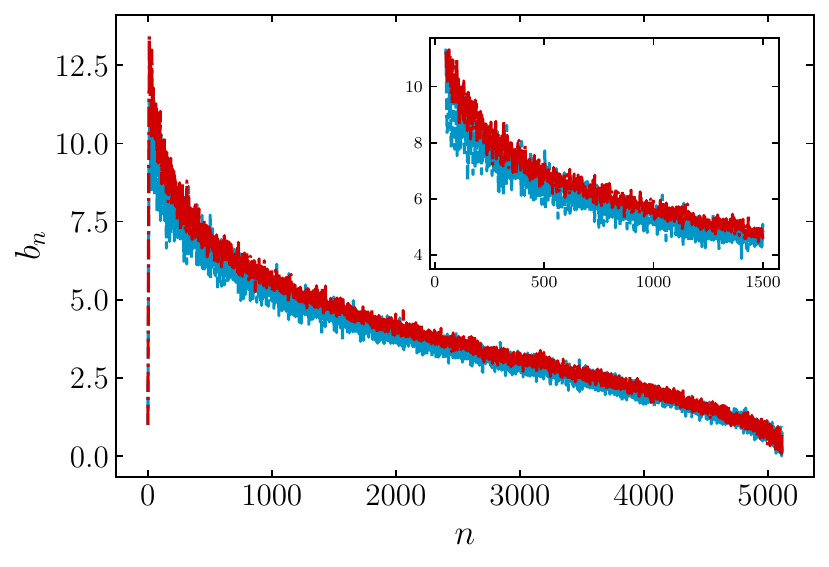}
     \caption{The complete lanczos-sequence is calculated for mixed-field TFIM for $L = 7$ spins in the $P = +1$ sector. The comaprison between the coefficients is shown for integrable and chaotic values with \textbf{Left}: $\alpha = 0.5$ \textbf{Right}: $\alpha =1$. The inset shows that the fluctuations for integrable parameter value (blue) are larger compared to chaotic parameter value (red). These larger fluctuation in large $b_n$ leads to suppression of K-complexity for integrable parameter values even for non-local model.}
     \label{fig:bn_MIXEDTFIM}
 \end{figure}

 \begin{figure}[h]
     \centering
     \includegraphics[width = 0.48\linewidth]{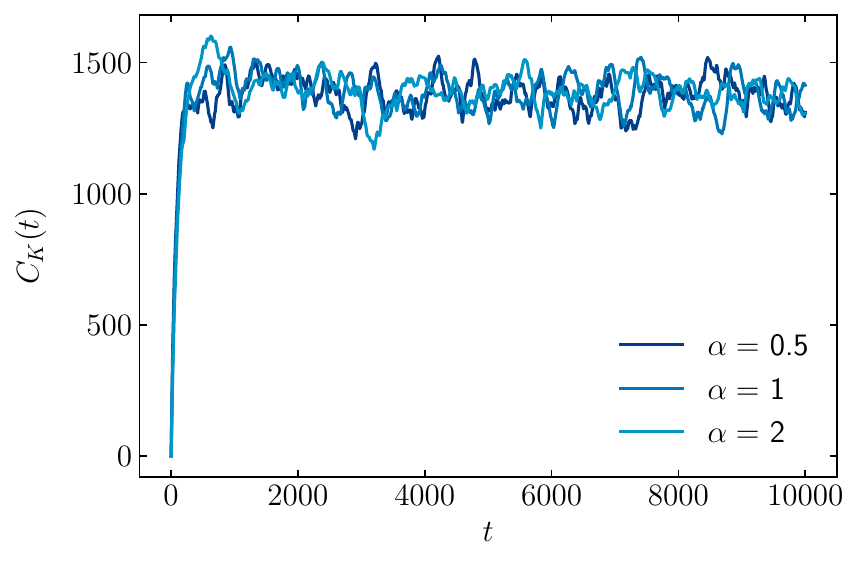}
     \includegraphics[width=0.48\linewidth]{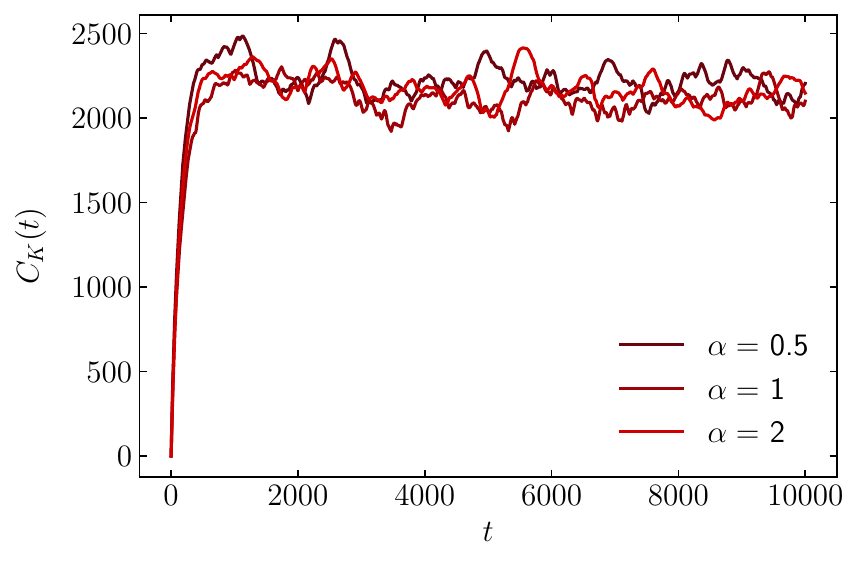}
     \caption{Late time Krylov complexity for the Transverse Mixed Field Ising Model Hamiltonian computed for $L = 7$ spins with open boundary conditions for fixed parity $P = +1$ in (a) Integrable (b) Chaotic limits with varying non-local exponent $\alpha$ values.}
     \label{fig:latetime_MIXEDTFIM}
 \end{figure}

The saturation value of the late-time Krylov Complexity is controlled by the amount of fluctuations in the Lanczos coefficients as reported in \cite{rabinoviciKrylovLocalizationSuppression2022, bhattacharyaKrylovComplexityOpen2023d}. Fluctuations are more for integrable parameter values and this phenomenon is dubbed as the Krylov localization. We notice that although non-local terms in the Hamiltonian increase the initial growth in Lanczos coefficients for both integrable and chaotic parameter values, the fluctuations in later Lanczos coefficients is still comparatively more for the integrable parameter values. This is shown in Figure \ref{fig:bn_MIXEDTFIM} for two choices of the non-local exponent $\alpha$ of the mixed TFIM model. As a result, the late time saturation value of Krylov complexity in Figure \ref{fig:latetime_MIXEDTFIM} distinctly differentiates the nature of the underlying integrable and chaotic theories even in the non-local models. In Figure \ref{fig:latetime_MIXEDTFIM}, the left figure shows the complexity saturation for various non-local $+$ integrable cases while the same is shown for non-local $+$ chaotic cases for various $\alpha$. While the late-time saturation value indeed distinguishes whether the underlying theory is integrable or chaotic, it remains insensitive to the magnitude of the non-local exponent $\alpha$. Nevertheless, as we noted earlier, the pre-scrambling growth of both Lanczos coefficients as well as the Krylov complexity serves as a clear indicator distinguishing between the non-local behaviors associated with different values of the exponent (Figure \ref{fig:lanczos_mixed_TFIM}).

	\subsection{Mixed field XXZ Hamiltonian}\label{subsec:mixedfieldXXZ}
	
In this section, we will introduce mixed-field XXZ Hamiltonian analogous to mixed-field TFIM. The XXZ Hamiltonian contains nearest-neighbour interaction terms which are used to describe the behavior of a system of interacting spin-1/2 particles in a magnetic field and the Hamiltonian can be written as,  
	\begin{equation}\label{eq:XXZ}
		H_\text{XXZ} = \frac{1}{4}\sum_{i=1}^{N-1} J\,(\sigma_i^x \sigma_{i+1}^x+\sigma_i^y \sigma_{i+1}^y) + \frac{J_{zz}}{4}\,\sigma_i^z \sigma_{i+1}^z\,.
	\end{equation}
The XXZ Hamiltonian commutes with total spin operator $M$ in the $z$-direction and is also invariant under reflection with respect to the edge of the chain i.e. under parity operator $P$. In previous studies \cite{Santos_2004,PhysRevE.84.016206,PhysRevB.80.125118,PhysRevB.98.235128,PhysRevB.102.075127, PhysRevX.10.041017,gubin_quantum_2012,rabinovici_krylov_2022}, it was shown that the integrability can be broken by addition local term   
\begin{equation}
    \begin{split}
        H_d  &= S_j^z\,.
    \end{split}
\end{equation}
The mixed field XXZ is an extension of the XXZ Hamiltonian where we add varying non-local strength analogous to the case of mixed field TFIM. 

	\begin{equation}
		\begin{split}
					\mathcal{H}^{(\alpha)}_{XXZ} &=  \frac{1}{4}\sum_{i=1}^{N-1} J^\alpha_{ij}\,(\sigma_i^x \sigma_{i+1}^x+\sigma_i^y \sigma_{i+1}^y)\\
					& \qquad \qquad + \frac{\tilde{J}^\alpha_{ij}}{4}\,\sigma_i^z \sigma_{i+1}^z
		\end{split}
	\end{equation}
	where $J^\alpha_{ij}$ and $\tilde{J}^\alpha_{ij}$ are interaction strength between spins at position $i$ and $j$. These follow powerlaw behavior, $1/\kappa \cdot J/|i-j|^\alpha$, and $1/\kappa \cdot J_{zz}/|i-j|^\alpha$ respectively. Note that this model reduces to local XXZ model in the limit $\alpha \rightarrow \infty$. We consider the following non-local interpolating Hamiltonian:
\begin{equation}\label{eq:Mixed_XXZ}
\mathcal{H}_\text{non-local} = 	\mathcal{H}^{(\alpha)}_{XXZ}  + \epsilon_d H_d\,.
\end{equation} 
In our study, we will consider $H_d = S^{z}_{(N+1)/2}$, and therefore, both the symmetry-breaking terms keep the symmetries intact. In further calculations, we fix $J = 1$, and $J_{zz} = 1.1$ for all cases.

	\begin{figure}[h]
		\centering
			\includegraphics[scale=0.29]{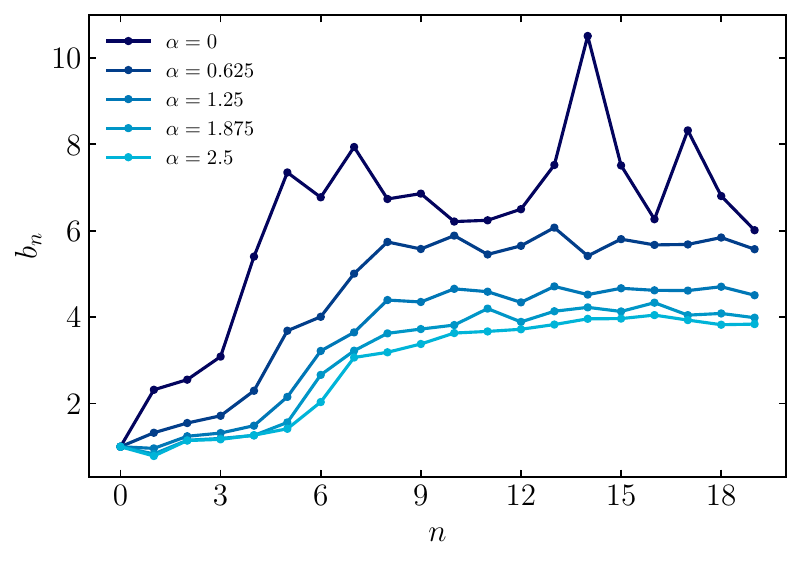}
			\includegraphics[scale=0.29]{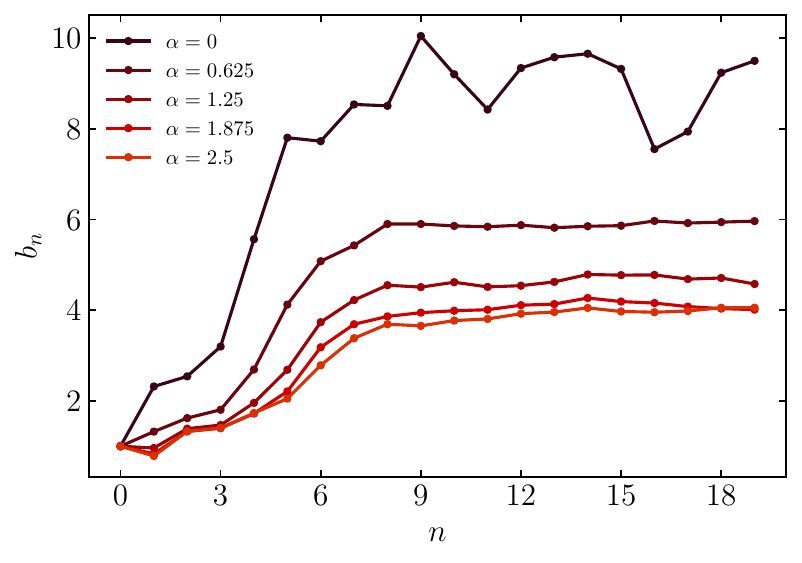}
		\caption{The Lanczos-sequence for the Transverse Mixed Field XXZ Model Hamiltonian in Eq.~\eqref{eq:mixed_TFIM} computed for $L = 12$ spins with open boundary conditions for fixed parity $P = +1$ and $M=5$ in (a) $\epsilon_d = 0$ (b) $\epsilon_d =0.5$ limits with varying non-local exponent $\alpha$ values. The initial operator is chosen to be $S^z_6 + S^z_7$. }
		\label{fig:lanczos_mixed_XXZ}
	\end{figure}

	\begin{figure}[h]
		\centering
			\includegraphics[scale=0.29]{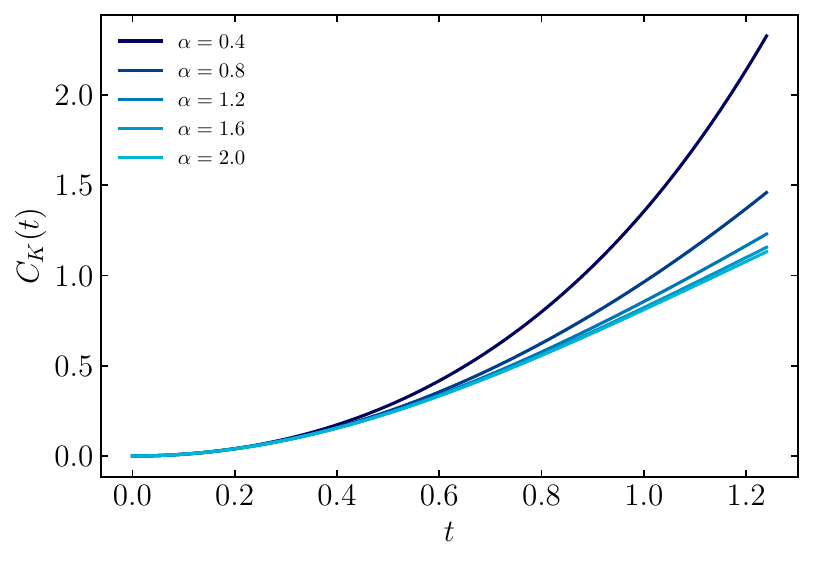}
			\includegraphics[scale=0.29]{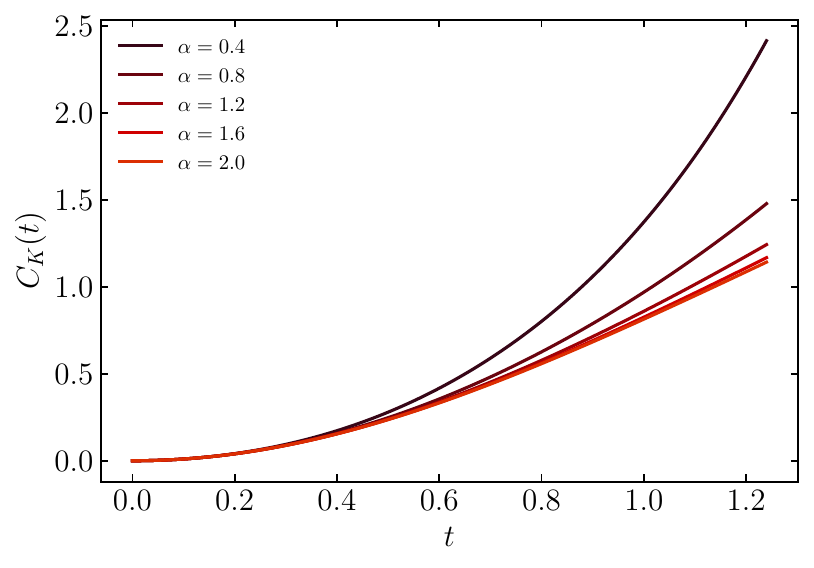}
		\caption{Krylov complexity for the Transverse Mixed Field XXZ Model Hamiltonian in Eq.~\eqref{eq:mixed_TFIM} computed for $L = 12$ spins with open boundary conditions for fixed parity $P = +1$ and $M=5$ in (a) $\epsilon_d = 0$ (b) $\epsilon_d =0.5$ limits with varying non-local exponent $\alpha$ values. }
		\label{fig:complexity_mixed_XXZ}
	\end{figure}

We study the initial Lanczos sequence for $L  = 12$ with varying exponent $\alpha$ for zero and non-zero value of $\epsilon_d$ (See Fig.~\ref{fig:lanczos_mixed_XXZ} and~\ref{fig:complexity_mixed_XXZ}). The initial operator $\mathcal{O}$ is chosen such that it respects the two symmetries of the model, i.e., parity and total spin in $z$-direction

\begin{equation}
\mathcal{O} = S^z_i + S^z_{N-i+1}
\end{equation}
where $i$ is chosen to be near the center of the chain. We find the increasing slope and saturation value in initial growth with the decrease in exponent $\alpha$ and, therefore increase in non-locality. These results are in agreement with the results presented in the main text for the mixed-field TFIM model and, therefore, reflect the universality of the result. We assert that the initial (pre-scrambling) growth rate of the Krylov complexity is always increased in the presence of non-locality.

	\section{Conclusions}\label{sec:Discussion}
	
\noindent To conclude, we have explored the behavior of K-complexity in non-local spin chains. The purpose of this study is to delineate what kind of novel features this quantity can capture in the presence of non-local couplings.  Over the course of this work, the K-complexity has emerged as a quantity of prime importance as it shows signs of non-locality in the system. Let us recount the major findings in the following.
	
	\subsection{Discussion of results}
	
	\noindent{\textbf{\textsf{Lanczos sequence}}} The saturation and initial growth of Lanczos coefficients is greater in non-local TFIM Hamiltonian relative to the local case, which reflects the faster information scrambling in non-local systems. The local chaotic dynamics, along with long-range non-local interactions, are sufficient to give rise to fast scrambling. This is because the Lanczos sequence for chaotic Hamiltonian with local interactions saturates at a comparatively smaller value than its non-local chaotic counterpart. The large saturation value of Lanczos coefficients are the result of non-local interactions in the system, but they can equivalently be said to be the result of fast scrambling as well. The fluctuation in the non-local cases is also suppressed, which is a result of integrability breaking due to non-local interaction.This results in a higher saturation value of Krylov complexity.\\

	\noindent{\textbf{\textsf{Krylov Complexity}}} Apart from common features of Krylov complexity, such as the transition from exponential growth to linear and the late time saturation, we find that the initial time growth rate is greater in non-local cases, indicating the onset of fast scrambling in the system. The higher growth rate in Krylov complexity for the non-local cases is a result of the initial faster growth of Lanczos coefficients in non-local cases due to the non-local interaction. Furthermore, since non-locality results in integrability breaking, we also find a large saturation value in Krylov complexity satisfying the bound in terms of the Krylov dimension. It is important to note that although the non-local interactions with the integrable local part of Hamiltonian show level statistics that of a chaotic model, it does not attain K-complexity saturation value similar to that of a chaotic case. In terms of associated Lanczos coefficients, the non-local integrable case has larger fluctuations as compared to the local chaotic case. It is noteworthy that even the introduction of the non-locality does not make the integrable model fully chaotic. On the other hand, when similar non-local terms are introduced in the otherwise local chaotic model, the late-time saturation value does not change much compared to the local chaotic model. However, the initial growth of complexity is faster in this case also. We reach similar conclusions through our study in the mixed field Ising and XXZ models. While the initial complexity growth acts as a probe of the degree of non-locality (different $\alpha$), the late-time saturation value only probes whether the underlying local theory ($\alpha > 2$) is integrable or not. \\

	\noindent In summary, our findings provide compelling evidence that operator growth, as characterized by K-complexity, exhibits a significantly greater magnitude in non-local systems when compared to their local counterparts. An increase in the non-locality parameter leads to a heightened scrambling rate in the system, as evident from both the Lanczos spectrum and the K-complexity profile. Furthermore, our observations indicate that the escalation of the non-locality parameter results in the breakdown of integrability and a transition towards chaotic behavior in the system. The integrability breaking due to non-local interaction is also evident from our study level statistics as well as spectral form factor (see Appendix~\ref{subsec:level_statistics} and \ref{subsec:SFF}).
	
	\subsection{Outlook}
	In future work, we would like to understand the relation between non-locality and fast scrambling better. This requires a careful investigation of scrambling time which is logarithmic in system size for fast scramblers, therefore requires investigation on system with large sizes.\footnote[4]{We leave this for future study due to the lack of resources to perform computation for higher dimensions.} It would be interesting find the degree of non-locality require for fast scramblers by studying model with power-law decaying interactions. In ref\,\cite{PhysRevA.102.022402}, the authors shown that the fast scrambling is prohibited in models with a generic all-to-all term with prefactor $\sim 1/N^\gamma$ if $\gamma > 1/2$. It would be interesting if this bound can be verified from operator growth using K-complexity.

Although all-to-all interactions are thought to be essential for fast scrambling\,\cite{belyansky2020minimal}, certain systems demonstrate fast scrambling even in the absence of such interactions\,\cite{ikedaCommentsMinimalModel2021}. These models consider spin chains with next-to-nearest neighbor interactions in a specified combination of some sectors. The spread of quantum information in similar models with mild non-locality are studied in literature\,\cite{ecclesSpeedingSpreadQuantum2021a}. These study show that although non-locality increases the rate of information spreading, but in lattice models these rates are suppressed.  An extension of this work may consider K-complexity in such mild non-local models.

	\begin{acknowledgments}
We wish to thank Sumilan Banerjee, Rathindra Nath Das, Bidyut Dey, Johanna Erdmenger, Mario Flory, Aaron Friedman, Michal Heller, Chethan Krishnan, Manas Kulkarni, Subroto Mukherjee, Pratik Nandy, and Aninda Sinha for various useful discussions and comments about this and related works. The work of AB is supported by the Polish National Science Centre (NCN) grant 2021/42/E/ST2/00234.
	\end{acknowledgments}

	\appendix
	
	\section{Appendices: Other probes of chaos: level statistics and spectral form factor}
	
	\subsection{Level Statistics}\label{subsec:level_statistics}
	
	In quantum chaos, Random Matrix Theory (RMT) is a powerful tool that accurately describes the spectral statistics of quantum systems whose classical counterparts exhibit chaotic behavior. For quantum Hamiltonians associated with integrable classical systems, the Berry-Tabor conjecture suggests that their level statistics follow a Poisson distribution\,\cite{10.1007/978-3-0348-8266-8_36}.
	
	However, for quantum Hamiltonians whose classical counterparts are chaotic, the Bohigas-Giannoni-Schmit conjecture proposes that their level statistics should fall into one of the three classical ensembles of RMT\,\cite{PhysRevLett.52.1}. These ensembles correspond to Hermitian random matrices with independently distributed entries as follows:
	
	\begin{itemize}
		\item Gaussian Orthogonal Ensemble (GOE): Corresponding to systems with real random variables as matrix entries.
		\item Gaussian Unitary Ensemble (GUE): Corresponding to systems with complex random variables as matrix entries.
		\item Gaussian Symplectic Ensemble (GSE): Corresponding to systems with quaternionic random variables as matrix entries.
	\end{itemize}
	These ensembles provide a robust framework for understanding the statistical behavior of quantum systems with chaotic classical dynamics. Consider $e_n$ be an ordered set of energy levels and the nearest-neighbor spacing, $s_n  = e_{n+1} - e_n$. Now, define the ratios $\Tilde{r}_n$ as 
	\begin{equation}
		\Tilde{r}_n  = \frac{\text{min}(s_n, s_{n-1})}{\text{max}(s_n,s_{n-1})} = \text{min}\left(r_n, \frac{1}{r_n}\right)
	\end{equation}
	where 
	\begin{equation}
		r_n = \frac{s_n}{s_{n-1}}\,.
	\end{equation}
	The distribution $P(r)$ for random matrix ensembles Wigner ensembles (GOE, GUE, and GSE) is shown to\,\cite{PhysRevLett.110.084101}
	
	\begin{equation}\label{eq:dist_wigner}
		P(r) = \frac{1}{Z_\beta} \frac{(r + r^2)^\beta}{(1+r+r^2)^{1+(3/2)\beta}} 
	\end{equation}
	with $Z_\beta$ the normalization constant. The Wigner ensembles are distinguished by their Dyson index ($\beta = 1,2$ and $4$ respectively) in distribution Eq.~\eqref{eq:dist_wigner}. For a Poissonian distribution of level-spacings, the distribution of $r$ is given by 
	\begin{equation}
		P(r) = \frac{1}{(1 + r)^2}
	\end{equation}
	The distribution $P(r)$ and $P(\Tilde{r})$ are related to one another by $P(\Tilde{r}) = 2P(r) \Theta(1-r)$. The mean value $\langle \Tilde{r} \rangle$ associated with different distribution (listed in Table \ref{table:mean_r}) can be used to distinguish between integrable and chaotic systems. 
	
	\begin{center}
		\begin{tabular}{ccccc} 
			\hline 
			Ensembles & Poisson & GOE & GUE & GSE \\
			\hline 
			& & & & \\
			$\langle \Tilde{r} \rangle$ & $2\ln 2-1$ & $4-2\sqrt{3}$ & $2 \frac{\sqrt{3}}{\pi} - \frac{1}{2}$ & $\frac{32}{15}\frac{\sqrt{3}}{\pi} - \frac{1}{2}$ \\ 
			& $\approx 0.38629$ & $\approx 0.53590$ & $\approx 0.60266$ & $\approx 0.67617$ \\
			\hline
		\end{tabular}
		\label{table:mean_r}
	\end{center}

	\begin{figure}[h]
		\centering
			\includegraphics[scale = 0.29]{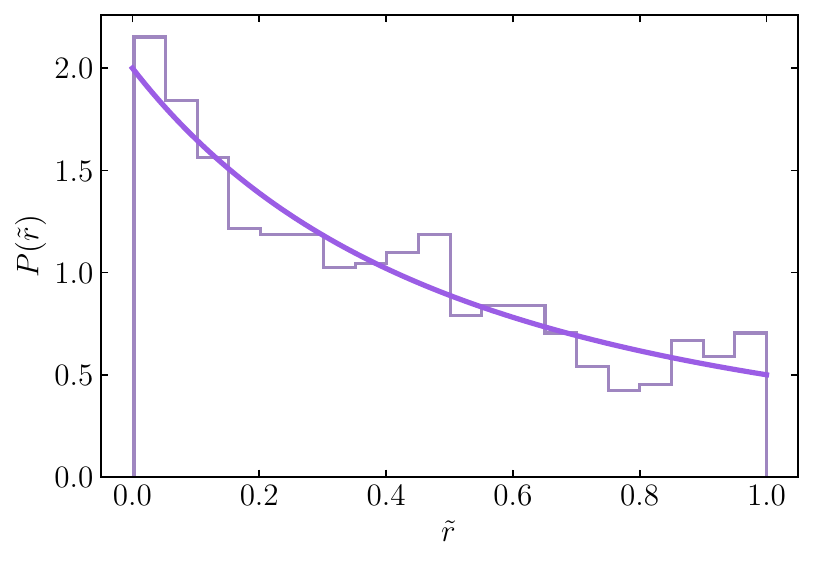}
			\includegraphics[scale = 0.29]{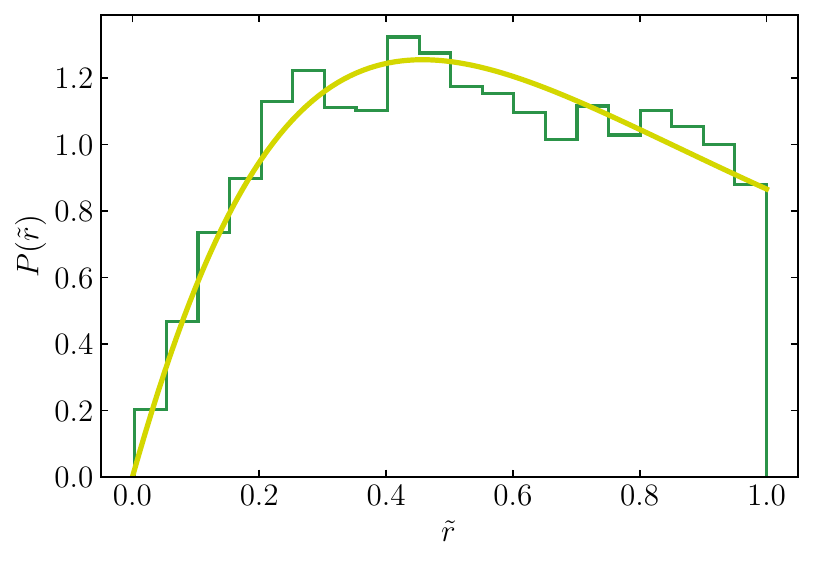}
		\caption{Probability distribution functions of the $\Tilde{r}$ statistics for fixed $Z$-reflection symmetry blocks and Parity block of the Hamiltonian \eqref{tfim}, computed for $L = 13$ spins in the $P = +1$, and $z = +1$ sector in (a) Integrable and (b) Chaotic limits. The results are obtained from averaging over 100 ensembles with $\epsilon$ drawn from normal distribution with mean $0$ and deviation $10^{-4}$. The mean level spacing ratio obtained for the integrable and chaotic limits are 0.38733 and 0.53433 respectively. }
		\label{fig:r_stat}
	\end{figure}

	\noindent{\textbf{\textsf{Results}}} In Fig.~\ref{fig:r_stat}, we showed the probability distribution functions of the $\Tilde{r}$ statistics of the local TFIM Hamiltonian \eqref{tfim} in integrable and chaotic regimes. We compare the results for both $P(\Tilde{r})$ and $\langle\Tilde{r}\rangle$ with the analytical results for Poisson and GOE symmetry class. The numerical result are in good agreement with those of analytical.\\

	\begin{figure}[h]
			\includegraphics[scale = 0.29]{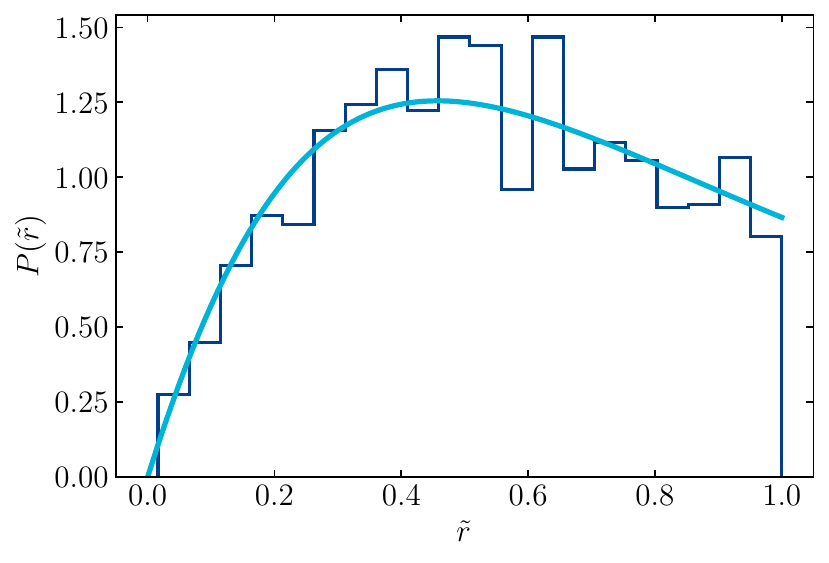}
			\includegraphics[scale = 0.29]{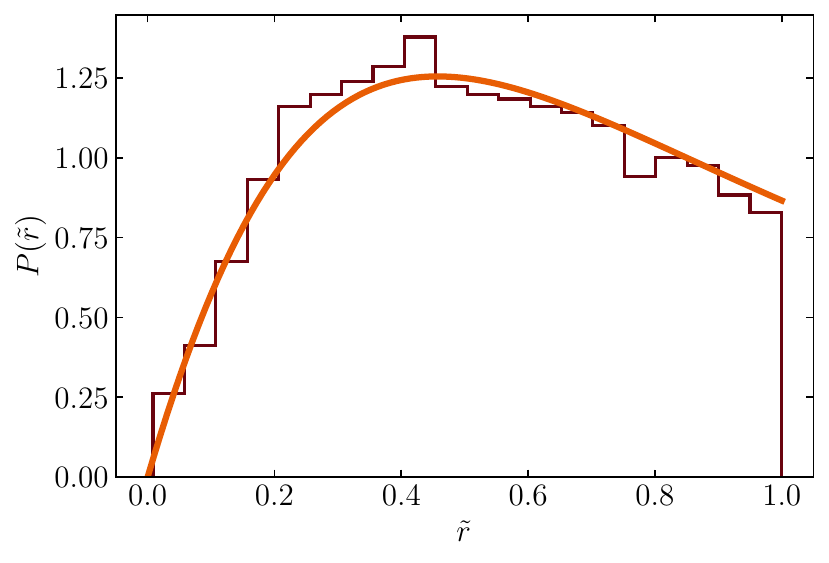}
		\caption{Probability distribution functions of the $\Tilde{r}$ statistics for fixed $Z$-reflection symmetry blocks and Parity block of the non-local Hamiltonian \eqref{tfim}, computed for $L = 13$ spins in the $P = +1$, and $z = +1$ sector and non-locality parameter $\gamma = 0.5$ in (a) Integrable and (b) Chaotic limits of local Hamiltonian. The results are obtained from averaging over 100 ensembles with $\epsilon$ drawn from a normal distribution with mean $0$ and deviation $10^{-4}$. The mean level spacing ratio obtained for integrable and chaotic limits are 0.53784 and 0.52738 respectively.}
		\label{fig:r_stat_NL}
	\end{figure}
	
	\noindent In Fig.~\ref{fig:r_stat_NL}, we showed the probability distribution function of the $\Tilde{r}$ statistics for non-local TFIM Hamiltonian (see Sec.~\ref{subsec:Non-local Transverse Field Ising Model}) computed for $L = 13$ spins. As reported in \cite{belyansky2020minimal}, we find that non-locality term changes the integrable model to chaotic for $|\gamma|\gtrsim 0.25$.

	\begin{figure}[h]
		\centering
			\includegraphics[scale = 0.29]{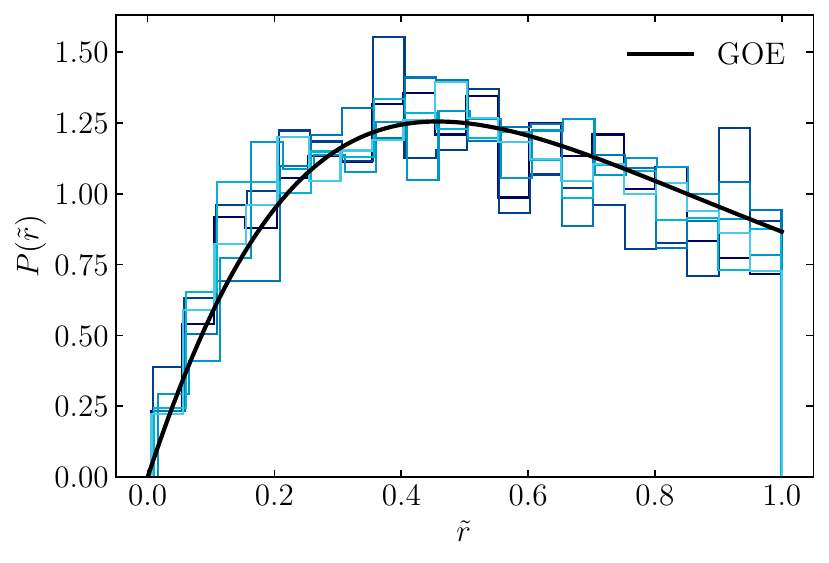}
			\includegraphics[scale = 0.29]{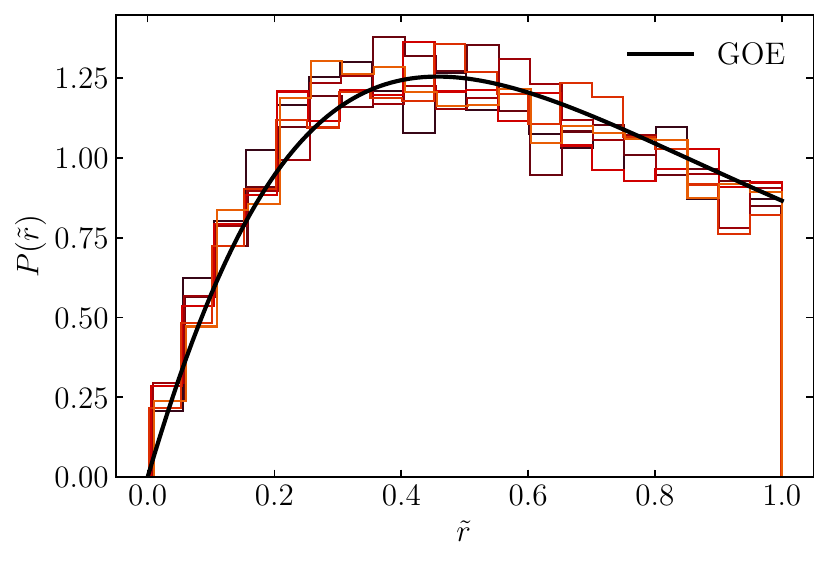}
		\caption{Probability distribution functions of the $\Tilde{r}$ statistics for Transverse mixed field Ising model for varying exponent $\alpha$. Refer Sec.~\ref{subsec:Transverse Mixed Field Ising} for parameter values.}
		\label{fig:rstat_Mixed_TFIM}
	\end{figure}

In Fig.~\ref{fig:rstat_Mixed_TFIM}, we shown level statistics distribution for Transverse mixed field Ising model discussed in Sec~\ref{subsec:Transverse Mixed Field Ising} for varying non-local exponent $\alpha$. It is clear from here that it is hard to distinguish these different degrees of non-locality from the level statistics, while the initial growth rate of K-complexity can quite clearly distinguish them. The reason is that in K-complexity, we can distinguish between different degrees of non-locality during the pre-scrambling regime. At the same time, the level statistics can capture only the details about the nature of the spectrum, and there is no way to make such a timewise distinction.

	\subsection{Spectral Form Factor}\label{subsec:SFF}
	
To compute the spectral form factor, we use the analytically continued partition function

\begin{equation}
    Z(\beta ,t) \equiv \text{Tr}\left( e^{-\beta H - iHt}\right)
\end{equation}
of the usual partition function $Z(\beta ) = \text{Tr}\left(e^{-\beta H}\right)$. The function $Z(\beta,t)$ is an erratic function of time and, therefore oscillates largely at late times. The fluctuations are studied by considering the normalized squared quantity, the so-called spectral form factor (SFF)\,\cite{cotlerBlackHolesRandom2017,PhysRevD.95.126008,winerSpectralFormFactor2022}
	
	\begin{equation}\label{eq:SFF1}
		\begin{split}
					g(\beta,t) &=  \left|\frac{Z(\beta,t)}{Z(\beta)}\right|^2 \\
					&= \frac{1}{Z(\beta)^2} \sum_{m,n} e^{-\beta (E_m+E_n)}e^{i(E_m - E_n)t}
		\end{split}
	\end{equation}
	where $E_n$ is eigenvalue of energy eigenstate $|n\rangle.$ The long-time average can be written as 
	
	\begin{equation}
		\lim_{T\rightarrow \infty} \frac{1}{T} dt \left| \frac{Z(\beta,t)}{Z(\beta)}\right|^2 = \frac{Z(2\beta)}{Z(\beta)^2}
	\end{equation}
	where we assumed the system Hamiltonian $H$ to be non-degenerate. Due to the erratic nature of $Z(\beta,t)$, usually the ensemble average over large Hamiltonians is considered. We define disorder-averaged analogs of SFF in Eq.~\eqref{eq:SFF1} as 
	
	\begin{align}
		g_\text{ann}(\beta,t) &= \frac{ \mathbb{E}[|Z(\beta,t)|^2]}{\mathbb{E}[|Z(\beta,0)|^2]} \\
		g_\text{que}(\beta,t) &= \mathbb{E} \left[ \left|\frac{Z(\beta,t)}{Z(\beta,0)}\right|^2 \right]
	\end{align}
 
which are called annealed SFF and quenched SFF, respectively. Throughout this paper, we use the notatin $\mathbb{E}[.]$ to denote the ensemble average. In this work, we will be working with annealed SFF along with $\beta = 0$, meaning that we are taking the disorder average separately in the numerator and denominator. We will further call $g_\text{ann}(\beta,0) = g(t)$, sometime this itself is referred to as the spectral form factor, which has information about the correlations of eigenvalues at different energy separations.
To understand the time profile of SFF, we split the $|\mathbb{E}[|Z(\beta,t)|^2]$ into two parts as 
	
	\begin{equation}
		\begin{split}
			&	|\mathbb{E} \text{Tr}(e^{-iH(\beta + it)})|^2  \\
				  & + \left(\mathbb{E}\left[|\text{Tr}(e^{-iH(\beta + it)})|^2 \right] - |\mathbb{E} \text{Tr}(e^{-iH(\beta + it)})|^2\right)
		\end{split}
	\end{equation}
	The first term is a disconnected part of the SFF, which comes solely from the average density of states. The second part is a connected part which contains the information on the correlation between energy levels. According to random matrix universality, an ensemble of quantum chaotic Hamilontians will generically have the same \textit{connected} SFF as the canonical Gaussian ensembles of RMT. The conjectured universal profile of the SFF of the GUE contains three distinct regimes \cite{winerSpectralFormFactor2022,Cipolloni_2023,brezinSpectralFormFactor1997}:
	
\begin{itemize}
    \item Initially the value drops quickly, through a region we call the \textit{slope}, to a minimum, which is called the \textit{dip}. The early time dip in profile comes from the disconnected part of the SFF therefore, its precise shape is non-universal. It is the result of loss of constructive interference in different terms of $\text{Tr} e^{-iH(\beta + it)}$ which acquire different phase factors as $t$ increases.
    \item After a dip, the values increase roughly linearly which is what we call the \textit{ramp}. The ramp is due to the repulsion between eigenvalues that are far apart in the spectrum, which is also well-known in quantum chaotic systems. Therefore a linear ramp is often taken as a defining signature of quantum chaos.
    \item The \textit{plateau}, which occurs at late times, results from the discreteness of the spectrum. 
\end{itemize}

	\begin{figure}[h]
		\centering
			\includegraphics[scale = 0.29]{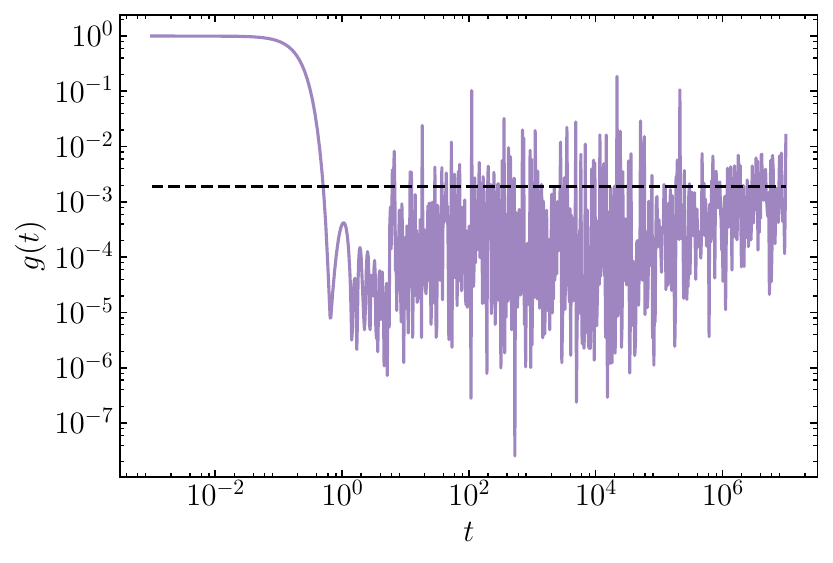}
			\includegraphics[scale = 0.29]{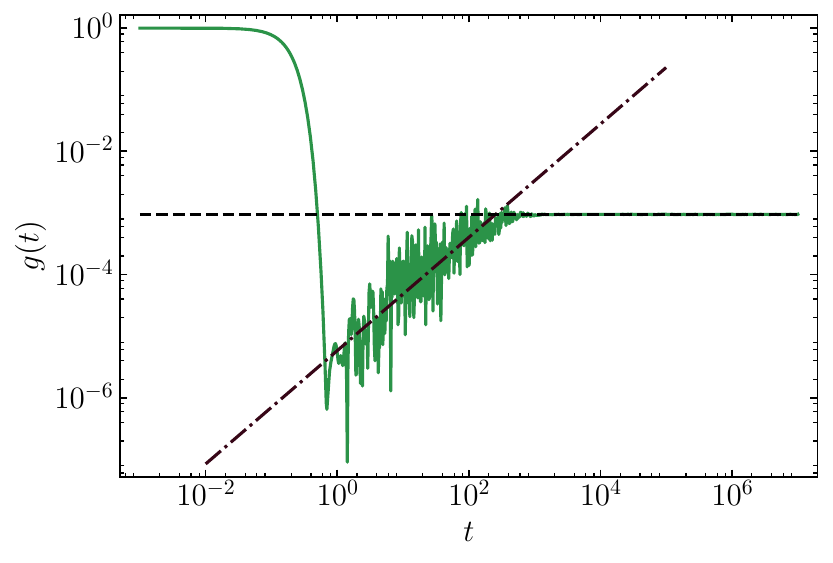}
		\caption{A log-log plot of TFIM Hamiltonian SFF $g(t;\beta = 0)$, plotted against time for $L = 11$ in (a) Integrable (b) Chaotic limits. The value at late times, which is equal to plateau height $g_p$, matches with $Z(2\beta)/Z(\beta)^2$ (shown in black dashed line). The average is taken over $50,000$ samples drawn from the random value of $\epsilon$ taken from a normal distribution with mean $\mu = 0$ and standard deviation $\sigma = 0.01$ in case of chaotic limit. The dash-dot line shows the linear fit in the ramp region, showing the linear increase. }
		\label{fig:SFF_TFIM_L}
	\end{figure}
	
The overall profile for GOE and GSE symmetry class also follows a slope-dip-ramp-plateau pattern apart from a few details, such as the shape of the ramp and the plateau. For example, the SFF of the GUE symmetry class has a sharp kink, while in the GOE symmetry class, the kink is smoothed\,\cite{Cipolloni_2023}.

	\begin{figure}[h]
		\centering
			\includegraphics[scale = 0.29]{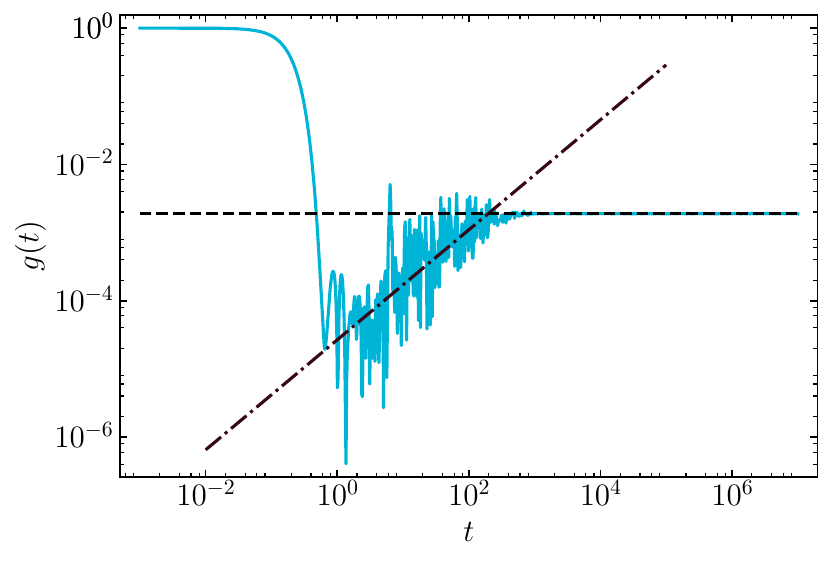}
			\includegraphics[scale = 0.29]{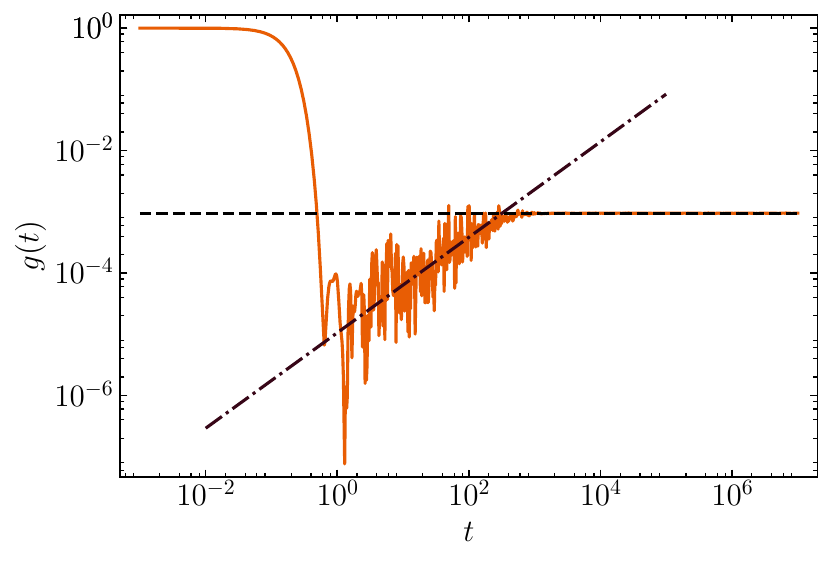}
		\caption{A log-log plot of non-local TFIM Hamiltonian SFF $g(t;\beta = 0)$, plotted against time for $L = 11$, and $\gamma = 0.5$ in (a) Integrable (b) Chaotic limits of local TFIM Hamiltonian. The value at late times, which is equal to plateau height $g_p$, matches with $Z(2\beta)/Z(\beta)^2$ (shown in black dashed line). The average is taken over $50,000$ samples drawn from the random value of $\epsilon$ taken from normal distribution with mean $\mu = 0$ and standard deviation $\sigma = 0.01$. The dash-dot line shows the linear fit in the ramp region, showing the linear increase.}
		\label{fig:SFF_TFIM_NL}
	\end{figure}

	\noindent
	\noindent{\textbf{\textsf{Results}}} In Fig.~\ref{fig:SFF_TFIM_L}, we showed the behavior of SFF of the TFIM Hamiltonian\,\eqref{tfim}, where an average is taken for the chaotic case. In the calculation of SFF for integrable cases, we didn't consider the disorder average since the model is no longer integrable with the addition of disorder. It is still a ``free-fermion model'' in the sense that it is bilinear in fermion operators. But it is not integrable anymore. While there are still an extensive number of conserved quantities, they are no longer quasilocal, so we expect to see chaotic behavior. In our calculation, we do find that the addition of disorder in integrable cases gives rise to a ramp in the SFF profile. In previous studies \cite{das_synthetic_2023}, it is shown that integrable systems such as rectangular billiards, SYK$_2$ model, and square-mod systems don't show ramps in their SFF profile. However, the emergence of a ramp-like feature has been demonstrated after averaging over the ensemble in the SYK$_2$ model. In Sec.~\ref{subsec:level_statistics}, we see that the model follows the statistics of the GOE symmetry class in the chaotic limit; therefore, we would expect such a profile to be the same as that of GOE symmetry class. In other words, in chaotic limit, we find slope-dip-ramp-plateau profile as evident from figure. \\

	\noindent In Fig. \ref{fig:SFF_TFIM_NL}, we show the SFF of non-local TFIM Hamiltonian for $L = 11$, and $\gamma = 0.5$ in the Integrable and Chaotic limits. We took the disordered average over 50,000 samples drawn from the random value of $\epsilon$. We find slope-dip-ramp-plateau patterns in both limits, integrable and chaotic, indicating that the non local interaction makes the Hamiltonian chaotic even when its local counterpart is in the integrable regime. It is also worth noting that there is no clear way to distinguish between the two non-local plots here. On the other hand, as described in the main text, by looking at the saturation values of the non-local K-complexity plots, we can still distinguish between which was integrable and which was not in the local version.

	\vfill
	
	\newpage
	\bibliographystyle{apsrev4-1}
	\bibliography{references} 
	
\end{document}